\documentclass[a4paper,12pt]{article}
\pdfoutput=1 

\usepackage{jheppub} 

\usepackage{amsmath, mathtools}
\usepackage{amsfonts}
\usepackage{amssymb}
\usepackage{accents}
\usepackage{color}
\usepackage{slashed}
\usepackage{comment}
\usepackage{hyperref}
\usepackage{booktabs}
\usepackage{braket}
\usepackage{cellspace}

\numberwithin{equation}{section}

\newcommand{\pri}[1]{\accentset{\prime}{#1}}

\newcommand{\AdS}[1]{\text{AdS}_{#1}}
\renewcommand{\S}{\text{S}}
\newcommand{\T}{\text{T}}
\newcommand{\AdSSSS}{\text{AdS}_3\times\text{S}^3\times\text{S}^3\times\text{S}^1}

\newcommand{\sph}[0]{\phi}
\newcommand{\sphh}[0]{\psi}
\newcommand{\Rads}[0]{R}

\title{The plane-wave limit of AdS$_{\text 3} \boldsymbol\times $S$^{\text 3} \boldsymbol\times$S$^{\text 3} \boldsymbol\times$S$^{\text 1}$}
\author{Andrea Dei,}
\author{Matthias R.~Gaberdiel, and Alessandro Sfondrini}

\affiliation{Institut f\"ur theoretische Physik, ETH Z\"urich\\ Wolfgang-Pauli-Stra{ss}e 27, 8093 Z\"urich, Switzerland}

\emailAdd{adei@itp.phys.ethz.ch}
\emailAdd{gaberdiel@itp.phys.ethz.ch}
\emailAdd{sfondria@itp.phys.ethz.ch}

\abstract{The plane-wave limit of ${\rm AdS}_3 \times {\rm S}^3 \times {\rm S}^3 \times {\rm S}^1$ is analysed for generic null-geodesics that are not necessarily BPS. For the case of pure NS-NS flux it is shown how the resulting spectrum can be reproduced as a suitable limit of the world-sheet description in terms of WZW models. Since supersymmetry is broken, most of the degeneracies are lifted, and thus the identification of states is quite unambiguous.}

\begin{document}
\maketitle
\flushbottom


\section{Introduction and conclusions}
Strings on $\AdS{3}$ have a rich dynamics. Even when restricting to the most supersymmetric backgrounds, which preserve 16 Killing spinors, the geometry is more varied than in higher dimensions. We have two geometries: $\AdS{3}\times\S^3\times\T^4$ and $\AdSSSS$. The latter is actually a one-parameter family of inequivalent geometries, which can be labelled by the ratio of the curvature radii of the two spheres. In the limit when one of the two spheres becomes flat, we find $\AdS{3}\times\S^3\times\mathbb{R}^3\times\S^1$ which is closely related to $\AdS{3}\times\S^3\times\T^4$ in the sector without winding and momentum on the flat coordinates. All these geometries can be supported by a mixture of Ramond-Ramond (R-R) and Neveu-Schwarz-Neveu-Schwarz (NS-NS) three-form fluxes, 
see, \textit{e.g.}~\cite{David:2002wn} for a review. 

Some of these backgrounds are directly accessible from a world-sheet perspective. In particular, for pure NS-NS backgrounds, the AdS$_3$ factor can be described by a Wess-Zumino-Witten (WZW) model based on $\mathfrak{sl}(2,\mathbb{R})$ 
\cite{Giveon:1998ns,Maldacena:2000hw, Pakman:2003cu,Israel:2003ry, Raju:2007uj,Ferreira:2017pgt,%
Giribet:2018ada,Gaberdiel:2018rqv} that is explicitly solvable. The situation for mixed (or pure R-R) flux is more complicated, but there exists the hybrid formalism of \cite{Berkovits:1999im} that gives an explicit description away from the pure WZW point, see also  \cite{Dolan:1999dc,Schomerus:2005bf,Gotz:2006qp,Saleur:2006tf,Troost:2011fd,Gaberdiel:2011vf,Gerigk:2012cq,KevinLorenz} for further developments. 
On the other hand, the pure R-R and mixed-flux backgrounds are known to be classically integrable fields theories \cite{Babichenko:2009dk,Sundin:2012gc,Cagnazzo:2012se}. This raises the hope of describing their quantum spectrum by the techniques that proved remarkably successful for strings on $\AdS{5}\times\S^5$ and $\AdS{4}\times\mathbb{C}\text{P}^3$, see refs.~\cite{Arutyunov:2009ga,Beisert:2010jr,Sfondrini:2014via} for reviews. Indeed an integrable world-sheet S-matrix was found for pure R-R backgrounds on $\AdSSSS$~\cite{Borsato:2012ud} and $\AdS{3}\times\S^4\times\T^4$~\cite{Borsato:2013qpa,Borsato:2014exa}, and later was generalized to mixed-flux backgrounds~\cite{Hoare:2013pma,Lloyd:2014bsa,Borsato:2015mma}. However, not all the dressing factors have been computed, and a closed set of equations for the finite-size spectrum is lacking.%
\footnote{A complication of these backgrounds is the presence of massless excitations, which makes the usual finite-size wrapping effects~\cite{Ambjorn:2005wa} much stronger here~\cite{Abbott:2015pps}.}

Some quantitative progress can be achieved by perturbatively quantizing the theory around a free point, \textit{i.e.}~by considering the string non-linear sigma model at large tension. It is well known that in the Penrose limit~\cite{Penrose1976, Blau:2001ne,Blau:2002dy,Berenstein:2002jq} we recover a free light-cone Hamiltonian, which for these backgrounds was already discussed in refs.~\cite{Berenstein:2002jq,Babichenko:2009dk, Rughoonauth:2012qd}. We can also compute perturbatively the corrections to these free theories; this was done for the purpose of computing the world-sheet S~matrix up to one~\cite{Sundin:2013ypa,Roiban:2014cia} and two loops~\cite{Sundin:2014ema,Sundin:2015uva}.
\smallskip

In this paper we study a family of inequivalent Penrose limits of $\AdSSSS$ in detail. 
As we will explain below (and as was observed already in refs.~\cite{Babichenko:2009dk,Rughoonauth:2012qd}), geodesics on $\AdS{3}\times\S^3\times\S^3$ are described by three parameters, one of which can be eliminated by the null condition. Of the resulting two-parameter family, only those having the same angular momenta along the two three-sphere are supersymmetric~\cite{Babichenko:2009dk};\footnote{In contrast, the null condition for a geodesic on $\AdS{3}\times\S^3\times\T^4$ that sits at a fixed point in $\T^4$ automatically implies that it is BPS.} however, the whole two parameter family yields a consistent Penrose limit --- though the spectrum of excitations around the ``plane-wave'' geometry does not need to be supersymmetric for generic choices of (non-BPS) geodesic.

Our main aim is to obtain this ``plane-wave'' spectrum for generic fluxes and geodesics. For pure NS-NS backgrounds we shall also compare our results with the WZW result, extending the analysis of \cite{Son:2003zv} to these more general cases. In the Green-Schwarz description of the string spectrum, states are obtained by acting with free oscillators (subject only to the level-matching constraint). This is to be compared to the WZW approach in which we need different ``spectrally flowed'' sectors in order to produce the whole spectrum.
While of course a matching is expected, it is nonetheless quite instructive to see how this comes about, and how the different ``flowed sectors'' are identified with states of the Green-Schwarz (GS) description.

Since our Penrose limits are generically taken around non-BPS geodesics, the resulting theories are typically non-supersymmetric. In particular, most of the degeneracies of the plane-wave spectrum are lifted, and the resulting spectrum is not just  fixed by supersymmetry. In the most general case, our light-cone Hamiltonian depends on \emph{four parameters}: two describe the original background ({\it i.e.}\ the relative curvature of the two three-spheres and the mixture of R-R/NS-NS fluxes) while two describe the geodesic chosen for the Penrose limit ({\it i.e.}\ how fast it is spinning and the relative speed on the two-three spheres). This wealth of parameters will allow for a precision-spectroscopy comparison of new proposals for computing the spectrum of strings on $\AdS{3}$ backgrounds, see {\it e.g.}\ \cite{Baggio:2018gct}.
\medskip

This paper is organised as  follows. In section~\ref{sec:Penrose-limit} we describe the different Penrose limits of $\AdSSSS$ with mixed background fluxes. 
In section~\ref{sec:the-green-schwarz-action} we derive the light-cone Hamiltonian for bosons and fermions in the Green-Schwarz formalism. After briefly reviewing the $\mathcal{N}=1$ WZW construction for this background in section~\ref{sec:the-wzw-spectrum}, we expand it in the near-plane wave limit in section~\ref{sec:plane-wave-limit}. Finally we compare these findings with the GS results in section~\ref{sec:comparison}. There are four appendices to which we have relegated a number of technical considerations.

\section{Penrose limits of \texorpdfstring{$\AdSSSS$}{AdS3xS3xS3xS1}}
\label{sec:Penrose-limit}

In this section we shall review how to obtain the plane-wave background from a Penrose limit~\cite{Penrose1976, Blau:2001ne,Blau:2002dy,Berenstein:2002jq} of AdS$_3$ geometries. We shall consider both the cases with Ramond-Ramond (R-R) and Neveu-Schwarz-Neveu-Schwarz (NS-NS) three-form fluxes. The resulting metric will take the form
\begin{equation} \label{eq:plane-wave-form-of-the-metric}
ds^2 = g_{\mu \nu} dx^\mu dx^\nu = 2 dx^+ dx^- - A_{ij} x_i x_j (dx^+)^2 + dx_i dx_i \ , 
\end{equation}
where $x^\pm$ are light-cone coordinates, while $A^{ij}$ will be a constant matrix (that will depend on our choice of background fluxes and of the geodesic). 

\subsection{The \texorpdfstring{$\AdSSSS$}{AdS3xS3xS3xS1} background}

Let us begin by writing the metric of  $\AdSSSS$ as 
\begin{equation} 
ds^2 = G_{\mu \nu} dX^\mu dX^\nu =  \Rads^2 \, ds^2_{\rm{AdS}_3} + R_1^2 \, ds^2_{\rm{S}^3_1} + R_2^2 \, ds^2_{\rm{S}^3_2} + ds^2_{\rm{S}^1}\ . 
\label{eq:ads_3xS^3xS^3xS^1-metric}
\end{equation}
In order for string theory to be critical,\footnote{Equivalently, this follows from demanding that the background preserves supersymmetry.} the radius~$R$ of $\AdS{3}$ and the radii of the two three-spheres $R_1$,  $R_2$ must satisfy
\begin{equation}
\dfrac{1}{\Rads^2} = \dfrac{1}{R_1^2} + \dfrac{1}{R_2^2}\ . 
\end{equation}
It is convenient to parameterize this relation as 
\begin{equation}\label{2.4}
\dfrac{\Rads^2}{R_1^2} = \alpha \equiv \cos^2 \varphi\ , \qquad \dfrac{\Rads^2}{R_2^2} = 1-\alpha \equiv \sin^2 \varphi\ .
\end{equation}
We shall use the conventions of \cite{Borsato:2015mma}, in which the metric on AdS$_3$ and S$^3$ is given by
\begin{equation}
\begin{aligned} ds^2_{{\rm AdS}_3} & = -  \left(\frac{4 + X_1^2 + X_2^2}{4 - X_1^2 - X_2^2} \right)^2 dt^2  + \left(\frac{4}{4 - X_1^2 - X_2^2} \right)^2 (dX_1^2 + dX_2^2)\ ,      \\
ds^2_{{\rm S}^3_1} & =  \left(\frac{4 - X_3^2 - X_4^2}{4 + X_3^2 + X_4^2} \right)^2 d \sph^2  + \left(\frac{4}{4 +  X_3^2 + X_4^2} \right)^2 (dX_3^2 + dX_4^2)\ ,       \\
ds^2_{{\rm S}^3_2} & =  \left(\frac{4 - X_5^2 - X_6^2}{4 + X_5^2 + X_6^2} \right)^2 d \sphh^2  +  \left(\frac{4}{4 +  X_5^2 + X_6^2} \right)^2 (dX_5^2 + dX_6^2)\ ,   \\
ds^2_{{\rm S}^1} & = dX_8 dX_8\ , 
\end{aligned}
\end{equation}
where $t, \sph$ and $\sphh$ are isometric coordinates. 
This background solves the supergravity equations and in fact has 16 Killing spinors when it is supported by a mixture of R-R and NS-NS 3-form fluxes\footnote{The explicit form of the Kalb-Ramond field $B$ is given in appendix~\ref{app:details-green-schwarz-action}.}
\begin{equation}
F = 2 \sqrt{1 - q^2} \Rads^2 \left( \rm{Vol}(AdS_3) + \dfrac{1}{\cos^2 \varphi}\rm{Vol}(S^3_1) + \dfrac{1}{\sin^2 \varphi} \rm{Vol}(S^3_2) \right),  
\label{eq:F-ads}
\end{equation}
and
\begin{equation}
H = dB = 2 q  \Rads^2 \left( \rm{Vol}(AdS_3) + \dfrac{1}{\cos^2 \varphi}\rm{Vol}(S^3_1) + \dfrac{1}{\sin^2 \varphi} \rm{Vol}(S^3_2) \right)\ .   
\label{eq:H-ads}
\end{equation}
Here $0 \leq q \leq 1$ describes the ratio between R-R and NS-NS flux, with $q=0$ corresponding to pure R-R and $q=1$ to pure NS-NS flux, respectively.  The 16 Killing spinors close on the superisometry algebra $\mathfrak{d}(2,1; \alpha)_L \oplus \mathfrak{d}(2,1; \alpha)_R$~\cite{Sevrin:1988ew}, where the labels $L$ and $R$ stand for left and right, and refer to the chirality in the dual $\text{CFT}_2$.
Each copy of $\mathfrak{d}(2,1; \alpha)$ has as a bosonic subalgebra $\mathfrak{su}(1,1) \oplus \mathfrak{su}(2)_1 \oplus \mathfrak{su}(2)_2 $, so that the global isometry is $\mathfrak{so}(2,2) \oplus \mathfrak{so}(4)_1 \oplus \mathfrak{so}(4)_2$. The $\mathfrak{d}(2,1;\alpha)_L$ algebra is characterized by the BPS bound~\cite{Gunaydin:1988re,deBoer:1999gea}
\begin{equation}
j \geq \cos^2 \varphi \ j_{1} + \sin^2 \varphi \ j_{2}\ , 
\label{eq:BPS-left-bound}
\end{equation}
where $j$ is the lowest weight of $\mathfrak{su}(1,1)_L$, while $j_1$ and $j_2$ are the spins of $\mathfrak{su}(2)_{1L} \oplus \mathfrak{su}(2)_{2L}$. A similar formula holds for $\mathfrak{d}(2,1;\alpha)_R$, where the corresponding quantum numbers are $(\tilde{\jmath}, \tilde{\jmath}_1, \tilde{\jmath}_2)$.

\subsubsection{The \texorpdfstring{$\AdS{3}\times\S^3\times\T^4$}{AdS3xS3xT4} limit}

As an aside we mention that in the limit $\varphi \to 0$, equation \eqref{eq:ads_3xS^3xS^3xS^1-metric} reduces to the metric of $\AdS{3} \times \S^3 \times \T^4$. Geometrically, this corresponds to one of the radii of the three-spheres going to infinity so that we are left with four flat coordinates. From the point of view of representation theory, this is captured by a suitable contraction of the algebra $\mathfrak{d}(2,1; \alpha)^2$ into $\mathfrak{psu}(1,1|2)^2$ for $\alpha \to 0$. Most of the formulae that we will derive below will admit a smooth $\alpha\to0$ limit, and thus our results also apply to the plane-wave limit of $\AdS{3}\times\S^3\times\T^4$.

\subsection{Null geodesics}
\label{subsec:null-geodesics}

We are interested in studying null geodesics of $\rm{AdS}_3 \times \rm{S}^3 \times \rm{S}^3 \times \rm{S}^1$. The geodesic equation%
\begin{equation}
\ddot X^\mu + \Gamma^{\mu}_{\; \; \nu \lambda} \dot X^\nu \dot X^\lambda = 0 
\end{equation}
is solved by a constant motion along the isometric directions $t, \sph, \sphh$. Let
\begin{equation}
t(\tau) = 2\pi\alpha'\dfrac{\Delta}{\Rads^2} \, \tau\ , \qquad \sph(\tau) = 2\pi\alpha'\dfrac{J_1}{R_1^2}\, \tau\ , \qquad \sphh(\tau) = 2\pi\alpha'\dfrac{J_2}{R_2^2} \, \tau\ , \qquad X_i(\tau) = 0\ , 
\label{eq:geodesic}
\end{equation}
where $\tau$ is the affine parameter.\footnote{In this equation we make explicit the string tension $1/(2\pi\alpha')$; the dimensionful parameter $\alpha'$ should not be confused with the dimensionless geometric parameter $\alpha$ of $\mathfrak{d}(2,1;\alpha)$. We will often suppress factors of $2\pi\alpha'$.} 
We have also chosen the normalization so that $\Delta, J_1$ and $J_2$ are the energy and angular momenta around the three-spheres, respectively; this follows from Noether's theorem, using the form of the Lagrangian 
\begin{equation}
\label{eq:LagrangianGeodesics}
L = \frac{1}{4\pi \alpha'}g_{\mu \nu} \dot X^\mu \dot X^\nu\ .
\end{equation}
The condition for this class of geodesics to be null is
\begin{equation}\label{2.12}
1 = \left( \cos \varphi \frac{J_1}{\Delta} \right)^2 + \left( \sin \varphi \frac{J_2}{\Delta} \right)^2 \ , 
\end{equation} 
which is solved by 
\begin{equation} 
J_1 = \Delta \dfrac{\cos \omega}{\cos \varphi}\ ,  \qquad J_2 = \Delta \dfrac{\sin \omega}{\sin \varphi}\ , 
\label{eq:null-condition-solution}
\end{equation}
where $\omega$ is a free parameter.%
\footnote{%
Since $J_1, J_2, \Delta\geq0$ we should always take the positive roots of $\cos^2\omega, \cos^2\varphi$, \textit{etc.}; to streamline our notation we drop the resulting absolute values and assume, without loss of generality, that $\cos \omega \cos \varphi\geq0$ and  $\sin \omega \sin \varphi\geq0$.}
The BPS bound of $\mathfrak{d}(2,1;\alpha)$ requires
\begin{equation}
\Delta - \cos^2 \varphi \ J_1 - \sin^2 \varphi \ J_2 \geq 0\ ,
\label{eq:BPS-bound}
\end{equation}
which holds true for any $\omega$,
\begin{equation}
\Delta - \cos^2 \varphi \ J_1 - \sin^2 \varphi \ J_2 = \Delta (1 - \cos(\varphi - \omega)) \geq 0 \ .
\end{equation}
We note that among the one-parameter family of geodesics \eqref{eq:null-condition-solution}, there is only one which saturates the BPS bound. This happens when $\omega=\varphi$ \cite{Babichenko:2009dk}, and the corresponding solutions preserve one quarter of the supersymmetry and have
\begin{equation}
\label{eq:protectedstatesJ}
\Delta = J_1 = J_2\ . 
\end{equation} 
Indeed \emph{all} protected states of $\AdSSSS$ satisfy eq.~\eqref{eq:protectedstatesJ}~\cite{Baggio:2017kza,Eberhardt:2017fsi}. Because of this, the geodesic in eq.~\eqref{eq:null-condition-solution} at generic~$\omega$ is a supergravity state that is typically not BPS. 

\subsection{The Penrose limit}
\label{subsec:the-penrose-limit}

We will now see how to obtain a plane-wave background~\eqref{eq:plane-wave-form-of-the-metric} from the Penrose limit around the geodesics of section~\ref{subsec:null-geodesics}. We introduce light-cone coordinates adapted to the geodesic given by~(\ref{eq:geodesic}) and (\ref{eq:null-condition-solution}) by defining $X^\pm$ and $X^7$ via 
\begin{equation}
\begin{pmatrix} t \\ \sph \\ \sphh \end{pmatrix} = \begin{pmatrix}
1 & -\tfrac{1}{2}  & 0 \\   \cos \varphi \cos \omega \ \  & \tfrac{1}{2}\, \cos \varphi \cos \omega  \ \ & - \cos \varphi \sin \omega \\  \sin \varphi \sin \omega \ \  & \tfrac{1}{2}\, \sin \varphi \sin \omega \ \ & \sin \varphi \cos \omega \end{pmatrix} \begin{pmatrix}
X^+ \\ X^- \\ X^7 \end{pmatrix}. 
\label{eq:change-of-coordinates-matrix}
\end{equation}
We also rescale the transverse coordinates (except for the $X^8$ coordinate on ${\rm S}^1$) by the radii of the respective spheres, {\it i.e.}\ we define 
\begin{equation}
X_{1,2} = \frac{x_{1,2}}{R}\ , \qquad X_{3,4} = \frac{x_{3,4}}{R_1} \ , \qquad X_{5,6} = \frac{x_{5,6}}{R_2} \ ,
\end{equation}
and set
\begin{equation}
X^+ = x^+\ , \qquad X^- = \frac{x^-}{\Rads^2}\ , \qquad X^7 = \frac{x^7}{\Rads}\ .  
\end{equation}
Expanding for large $\Rads$, the metric in \eqref{eq:ads_3xS^3xS^3xS^1-metric} reduces to the form~\eqref{eq:plane-wave-form-of-the-metric}, \textit{i.e.}
\begin{equation}
ds^2 = 2 dx^+ dx^- - A_{ij} x_i x_j (dx^+)^2 + dx_i dx_i\ , \qquad i, j = 1, \dots 8\ ,
\end{equation}
where $A_{ij}$ is the mass-squared matrix for the eight transverse coordinates,%
\footnote{The superscript $b$ in the masses $\mu_i^b$ indicates that these will be the masses of the bosonic string excitations. Later we will also introduce Green-Schwarz fermions with mass $\mu_i^f$.}  
\begin{equation}
A_{ij} = \text{diag} \big( \mu_1^b,\mu_1^b,\ \mu_2^b,\mu_2^b,\ \mu_3^b,\mu_3^b,\ \mu_4^b,\mu_4^b \big)^2\ , 
\end{equation}
with
\begin{equation}
\mu_1^b=1\ ,\qquad
\mu_2^b=\cos \varphi \cos \omega\ ,\qquad
\mu_3^b=\sin \varphi \sin \omega\ ,\qquad
\mu_4^b=0\ .
\label{eq:boson-masses-GS}
\end{equation}
Notice that we have \emph{two} directions with $A_{ii}=0$, which will result in massless modes in the Green-Schwarz string formulation \cite{Babichenko:2009dk}. One is the flat direction corresponding to the $\S^1$ factor, while the other is the linear combination of $(\phi,\psi)$ on $\S^3\times\S^3$ that is orthogonal to the light-cone geodesic. The 3-form fluxes \eqref{eq:F-ads} and \eqref{eq:H-ads} reduce to 
\begin{multline}
F = 2 \sqrt{1 - q^2} \left(dx^+ \wedge dx_1 \wedge dx_2 + (\cos \varphi \cos \omega)\, dx^+ \wedge dx_3 \wedge dx_4 \right. \\
+ \left. (\sin \varphi \sin \omega)\, dx^+ \wedge dx_5 \wedge dx_6 \right)\ , 
\end{multline}
\begin{multline}
H = 2 q \left(dx^+ \wedge dx_1 \wedge dx_2 + (\cos \varphi \cos \omega)\, dx^+ \wedge dx_3 \wedge dx_4 \right. \\
+ \left. (\sin \varphi \sin \omega)\, dx^+ \wedge dx_5 \wedge dx_6 \right)\ .  
\end{multline}
In the following, $x^+$ and $x^-$ will play the role of light-cone coordinates. 
The conserved charges associated to the light-cone coordinates follow from the change of coordinates and eq.~\eqref{eq:null-condition-solution}\footnote{The sign in the definition of $\Delta$ (relative to $J_1$ and $J_2$) is a consequence of the signature of Minkowski space.}
\begin{align}
P_+=\int d\sigma\,p_+ &=  - \big(\Delta - (\cos \varphi \cos \omega)\, J_1 - (\sin \varphi \sin \omega) \, J_2 \big)\ ,
\label{P+} \\
P_-=\int d\sigma\,p_- &= \dfrac{1}{2\Rads^2} \big(\Delta + (\cos \varphi \cos \omega) \, J_1 + (\sin \varphi \sin \omega) \, J_2\big)\ .
\label{eq:p_+-p_-} 
\end{align}
The momentum $P_+$ plays a special role: it defines the \emph{light-cone Hamiltonian}~$H$, whose density~$\mathcal{H}$ is given by 
\begin{equation}
\mathcal{H} = -p_+\,,\qquad
H=\int d\sigma\, \mathcal{H}\,. 
\label{eq:H=-p_+}
\end{equation} 
On the other hand, using \eqref{eq:null-condition-solution} and keeping track of the factors of $2\pi\alpha'$, the momentum $P_-$ reduces to 
\begin{equation}
P_- = 2\pi\alpha'\,\dfrac{ \Delta}{\Rads^2}\,,
\label{eq:p_-propto-delta}
\end{equation} 
which parameterizes how fast the string is spinning along the chosen geodesic, \textit{cf.}\ eq.~\eqref{eq:geodesic}, where small (large) values of $P_-$ correspond to \emph{slowly} (\emph{rapidly}) \emph{spinning} geodesics. Since we are interested in configurations with finite momenta, we will from now on assume that $\Delta$, $J_1$ and $J_2$ scale as $\Rads^2$ in the Penrose limit, but in such a way that $H$ remains finite and $P_-$ is a free parameter. One can easily verify that this is consistent with \eqref{eq:null-condition-solution}; in fact, if $J_1$, $J_2$ are as in \eqref{eq:null-condition-solution}, $\mathcal{H} = p_+ = 0$, as $p_+$ is the momentum transverse to the geodesic. Note that the ground-state energy in light-cone gauge will receive quantum corrections, unless the ground state is BPS. 

\subsubsection{Penrose limit of \texorpdfstring{$\AdS{3}\times\S^3\times\T^4$}{AdS3xS3xT4}}

Again, as an aside, we note that for $\AdS{3}\times\S^3\times\T^4$, {\it i.e.}\ for $\varphi\rightarrow 0$, the null condition (\ref{2.12}) for the geodesics is simply that $J_1 = \Delta$. The null geodesic then runs along $\AdS{3}\times\S^3$, and it is automatically BPS. In fact, it preserves one half, rather than one quarter, of the Killing spinors. This leads to the same metric as in eq.~\eqref{eq:plane-wave-form-of-the-metric}, where the masses in the matrix $A_{ij}$ now take the values
\begin{equation}
\mu_1^b=\mu_2^b=1\,,\qquad
\mu_3^b=\mu_4^b=0\,.
\end{equation}

\section{The Green-Schwarz action}
\label{sec:the-green-schwarz-action}

In this section we will outline the computation of the Green-Schwarz action for strings moving in the plane-wave background described in section~\ref{sec:Penrose-limit}. We shall study the case of generic (mixed) flux, parametrised by $0\leq q \leq 1$, see the discussion below eq.~(\ref{eq:H-ads}). 

\subsection{The bosonic action}
\label{subsec:the-bosonic-action}
The bosonic part of the action is%
\footnote{Here we suppressed a factor of $1/(2\pi\alpha')$ in front of the action.}
\begin{equation}
S_B = -\dfrac{1}{2} \displaystyle\int d \sigma d \tau \left( \gamma^{\alpha \beta} g_{MN} \partial_\alpha x^M \partial_\beta x^N + \epsilon^{\alpha \beta} B_{MN}\partial_\alpha x^M \partial_\beta x^N \right)\ , 
\label{eq:bosonic-action}
\end{equation} 
where the determinant of the world-sheet metric $\gamma^{\alpha \beta}$ has been set to $-1$. Notice that for point-like strings, this action reduces to~\eqref{eq:LagrangianGeodesics}. To obtain the Hamiltonian it is convenient to work directly in first order formalism, introducing the conjugate momenta
\begin{equation}
p_M := \dfrac{\delta S_B}{\delta \dot{x}^M} = - \gamma^{\tau \beta} g_{MN} \partial_\beta x^N - B_{MN} \pri{x}^N\ , 
\label{eq:conjugate-momenta}
\end{equation}
where we denote with a prime and a dot the derivative with respect to the world-sheet coordinates $\sigma$ and $\tau$, respectively. The action can be rewritten in the form~\cite{Arutyunov:2009ga,Sfondrini:2014via, Lloyd:2014bsa,Borsato:2015mma}
\begin{equation}
S_B = \displaystyle\int d \sigma d \tau \left( p_M \dot{x}^M + \dfrac{\gamma^{\tau \sigma}}{\gamma^{\tau \tau}}C_1 + \dfrac{1}{2 \gamma^{\tau \tau}}C_2 \right)\ , 
\end{equation}
with 
\begin{equation}
\begin{gathered}
    C_1 = p_M \pri{x}^M \ , \\[4pt]
    C_2 = g^{MN} p_M p_N + g_{MN} \pri{x}^M \pri{x}^N + 2g^{MN} B_{NK} p_M \pri{x}^K + g^{MN} B_{MK} B_{NL} \pri{x}^K \pri{x}^L\  .
\end{gathered}
\label{eq:virasoro-constraints}
\end{equation}
The Virasoro constraints amount to the equations of motion for the auxiliary fields $\gamma^{\alpha \beta}$. For the bosonic action they are equivalent to setting $C_1 = 0$ and $C_2 = 0$. 

\subsection{The fermionic action}

The Green-Schwarz action for $\AdS{3}$ backgrounds can be written explicitly up to quartic order in the fermions~\cite{Grisaru:1985fv, Cvetic:1999zs, Wulff:2013kga}, and it reads
\begin{equation}
S = \displaystyle\int d \sigma d \tau \mathcal{L} =  \displaystyle\int d \sigma d \tau\left( \mathcal{L}_B + \mathcal{L}_{kin} + \mathcal{L}_{WZ} \right)\ , 
\label{eq:GS-action}
\end{equation}
where $S_B = \int \mathcal{L}_B$ has been defined in subsection \ref{subsec:the-bosonic-action}, and
\begin{equation}
\mathcal{L}_{kin} = -i \gamma^{\alpha \beta} \bar{\tilde{\theta}}_I \slashed{E}_\alpha \left( \delta^{IJ} D_\beta + \dfrac{1}{48} \sigma_3^{IJ} \slashed{F} \slashed{E}_\beta + \dfrac{1}{8} \sigma_1^{IJ}\slashed{H}_\beta \right) \tilde{\theta}_J\ ,
\label{eq:GS-L-kin} 
\end{equation}
\begin{equation}
\mathcal{L}_{WZ} = +i \epsilon^{\alpha \beta} \bar{\tilde{\theta}}_I \sigma_1^{IJ} \slashed{E}_\alpha \left( \delta^{JK} D_\beta + \dfrac{1}{48} \sigma_3^{JK} \slashed{F} \slashed{E}_\beta + \dfrac{1}{8} \sigma_1^{JK}\slashed{H}_\beta \right) \tilde{\theta}_K\ .
\label{eq:GS-L-WZ}
\end{equation}
Here we have only kept the terms that are quadratic in the fermions as the higher order terms will not be important in the plane-wave limit.
The indices $\alpha, \beta \in \{\tau, \sigma \}$ are world-sheet indices, while $I, J, K \in \{1, 2\}$. For the vielbeins $E_\mu^{\ A}$ we have 
\begin{equation}
E_\mu^{\ A} E_\nu^{\ B} \eta_{AB} = g_{\mu \nu}\ , \qquad \slashed{E}_\alpha = \partial_{\alpha} X^\mu \slashed{E}_\mu\ , \qquad \slashed{E}_\mu = \displaystyle\sum_{A = 0}^9 E_\mu^{\ A} \eta_{AB} \Gamma^B\ ,  
\end{equation}
while the fluxes are defined by 
\begin{equation}
\slashed{F} = F_{ABC} \Gamma^{ABC}\ , \qquad \slashed{H}_\alpha = \partial_{\alpha}  X^\mu \slashed{H}_\mu \ , \qquad \slashed{H}_\mu = H_{\mu A B}\Gamma^{AB}\ . 
\end{equation}
More technical details about \eqref{eq:GS-L-kin} and \eqref{eq:GS-L-WZ} can be found in appendix~\ref{app:details-green-schwarz-action}.

\subsection{Gauge fixing and kappa-symmetry fixing}

The gamma matrices associated to the light-cone directions $x^+$ and $x^-$ are
\begin{equation}
\begin{aligned}
\Gamma^+ &= \dfrac{1}{2} \left(\cos \omega \ \Gamma^\sph + \sin \omega \ \Gamma^\sphh + \Gamma^t \right)\ , \\
\Gamma^- & = \left(\cos \omega \ \Gamma^\sph + \sin \omega \ \Gamma^\sphh - \Gamma^t \right)\ , 
\end{aligned}
\label{eq:gamma+-}
\end{equation}
where our gamma matrix conventions are spelled out in appendix~\ref{app:conventions}. We note that, despite the involved form of the light-cone directions, $\Gamma^\pm$ still satisfy the familiar identities
\begin{equation}
(\Gamma^\pm)^2 = 0\ , \qquad \tfrac{1}{2}(\Gamma^+ \Gamma^- + \Gamma^- \Gamma^+)  = {\bf 1} \ . 
\label{eq:gamma+-identities}
\end{equation}
We fix light-cone gauge and kappa symmetry by setting
\begin{equation}
x^+ = \kappa\,\tau\ , \qquad p_- = \kappa'\ , \qquad \Gamma^+ \theta_I = 0\ ,
\end{equation}
where $\kappa$ and $\kappa'$ are two positive constants.
This is called \emph{uniform} light-cone gauge because the momentum density~$p_-$ along the string is constant~\cite{Arutyunov:2004yx,Arutyunov:2005hd, Arutyunov:2006gs}.  We should mention that the choice of $p_-$ determines the size~$r$ of the world-sheet after gauge-fixing. In particular we have, in the Penrose limit,
\begin{equation}
2\pi\alpha'\,\frac{\Delta}{R^2}=P_{-} = \int_0^r d\sigma\, p_- = r\,p_-\ ,
\end{equation}
where we have reinstated the factor of $2\pi\alpha'$ for future convenience.
In what follows we find it convenient to set
\begin{equation}
x^+ = \tau\ , \qquad p_- = 1\ , \qquad \Gamma^+ \theta_I = 0\ ,
\label{eq:gauge-fixing}
\end{equation}
so that the world-sheet length equals
\begin{equation}
r= P_- = 2\pi\alpha'\,\frac{\Delta}{R^2}\ .
\label{P_-=r}
\end{equation}
 As described in appendix~\ref{app:Gauge-fixed-Green-Schwarz-action-in-plane-wave-background}, eq.~\eqref{eq:GS-action} simplifies drastically for the plane-wave background described in section \ref{sec:Penrose-limit}. The resulting world-sheet Hamiltonian is quadratic in the fields,
\begin{equation}
\mathcal{H} = \mathcal{H}^b + \mathcal{H}^f, 
\end{equation}
where 
\begin{align}
\mathcal{H}^b  = & \frac{1}{2} \left( p_i p_i + \pri x_i \pri x_i + A_{ij} x_i x_j \right) + q \mu_1^b(x_2 \pri x_1 - x_1 \pri x_2)  \nonumber \\
& + q \mu_2^b (x_4 \pri x_3 - x_3 \pri x_4) + q \mu_3^b (x_6 \pri x_5 - x_5 \pri x_6) \ ,   
\label{eq:bosonic-hamiltonian} \\
\mathcal{H}^f = & \sum_{j = 1}^4 \bigl[i \bar{\theta}_{1j} ( i \tilde q \pri{\theta}_{2j} - q \pri{\theta}_{1j} ) 
- i \bar{\theta}_{2j} (  i \tilde q \pri{\theta}_{1j} - q \pri{\theta}_{2j} ) + \mu_j^f ( \bar{\theta}_{1j} \theta_{1j}  - \bar{\theta}_{2j} \theta_{2j} ) \bigr]\ , 
\label{eq:fermionic-hamiltonian}
\end{align}
and
\begin{equation}
\begin{aligned}
\mu_1^f & = \dfrac{1 + \cos(\phi-\omega)}{2}\ , \quad  & \quad  \mu_2^f & = \dfrac{1 + \cos(\phi+\omega)}{2}\ ,  \\
 \mu_3^f & = \dfrac{1 - \cos(\phi+\omega)}{2}\ , \quad & \quad \mu_4^f & =\dfrac{1 - \cos(\phi-\omega)}{2}\ . 
\end{aligned}
\label{eq:fermion-masses-GS}
\end{equation}
The parameter $q$ labels again the ratio of background fluxes. We should mention that 
the fermionic masses $\mu_1^f, \dots, \mu_4^f$ differ in general from the bosonic masses $\mu_1^b, \dots, \mu_4^b$ defined in \eqref{eq:boson-masses-GS}, and coincide only for BPS geodesics, {\it i.e.}\ for $\omega = \phi$.  

\subsection{Mode expansion}
\label{subsec:mode-expansion-and-normal-ordering}

The eight real coordinates $x_1, \dots , x_8$ enjoy an $\mathfrak{so}(2)^{\oplus4}$ symmetry. The  Hamiltonian \eqref{eq:bosonic-hamiltonian} can be diagonalised by introducing complex coordinates for each $\mathfrak{so}(2)$ doublet, yielding four complex bosons. We introduce for each complex boson a pair of creation and annihilation operators $a_n^{i, \pm \dagger}$ and $a_n^{i, \pm}$ with $i = 1, \dots, 4$ and%
\footnote{The explicit mode-expansion of the complex bosons can be found in appendix~\ref{app:Gauge-fixed-Green-Schwarz-action-in-plane-wave-background}.}
\begin{equation}
\left[ a_n^{i,\pm} , a_m^{i, \pm \dagger} \right]  = \delta_{nm}\ , \qquad \left[ a_n^{i,\pm} , a_m^{i, \mp \dagger} \right]  = 0\ . 
\label{eq:bosonic-commutation-relations}
\end{equation} 
Then the Hamiltonian takes the form
\begin{equation}
\begin{aligned}
H^b = \displaystyle\int_0^{r} d \sigma \mathcal{H}^b = & \displaystyle\sum_{i=1}^4  \displaystyle\sum_{n \in \mathbb{Z}} \left[ \omega_n^+(\mu_i^b) a_n^{i,+ \dagger} a_n^{i,+} + \omega_n^-(\mu_i^b) a_n^{i,- \dagger} a_n^{i,-}   \right] \\
& + \displaystyle\sum_{i=1}^4 \Big[|\mu_i^b| + \displaystyle\sum_{n > 0} \bigl( \omega_n^+(\mu_i^b) + \omega_n^-(\mu_i^b)\bigr) \Big] \ , 
\end{aligned}
\label{eq:bosonic-hamiltonian-modes}
\end{equation}
where the terms in the second line arise from normal ordering. Here, $\mu_i^b$ are the bosonic masses  defined in eq.~\eqref{eq:boson-masses-GS}, and the dispersion relation is
\begin{equation}
\label{eq:dispersion-relation}
\omega_n^{\pm} (\mu) = \sqrt{\mu^2 \pm 2 q \mu \frac{2\pi\, n}{r} + \frac{4\pi^2n^2}{r^2}}\ . 
\end{equation}
We can think of the quantity $p = \frac{2\pi n}{r}$ as the world-sheet momentum~$p$, subject to the free-particle quantization condition $\exp(i p r)=1$; thus we can also write the dispersion relation as 
\begin{equation}\label{3.24}
\omega_p^{\pm} (\mu) = \sqrt{\mu^2 \pm 2 q \mu p + p^2}\ . 
\end{equation}

The fermionic Hamiltonian \eqref{eq:fermionic-hamiltonian} can be diagonalised introducing creation and annihilation operators $b_n^{i,\pm \dagger}$ and $b_n^{i,\pm}$ with $i=1, \dots, 4$ and
\begin{equation}
\{b_n^{i, \pm \dagger}, b_m^{j, \pm} \} = \delta^{ij} \delta_{mn}\ , \qquad \{b_n^{i, \pm \dagger}, b_m^{i,\mp} \} = 0\ , 
\label{eq:fermionic-commutation-relations}
\end{equation}
where the explicit mode expansion is again given in appendix~\ref{app:Gauge-fixed-Green-Schwarz-action-in-plane-wave-background}.
Then eq.~\eqref{eq:fermionic-hamiltonian} reduces to 
\begin{equation}
\begin{aligned}
H^f = \displaystyle\int_0^{r} d \sigma \mathcal{H}^f = & \sum_{i=1}^4 \displaystyle\sum_{n \in \mathbb{Z}} \left[ \omega_n^+(\mu_i^f) b_n^{i,+ \dagger} b_n^{i,+} + \omega_n^-(\mu_i^f) b_n^{i, - \dagger} b_n^{i,-}  \right] \\
& - \displaystyle\sum_{i=1}^4 \Big[ |\mu_i^f| +\sum_{n > 0} \bigl(\omega_n^+(\mu_i^f) + \omega_n^-(\mu_i^f) \bigr) \Big] \ , 
\end{aligned}
\label{eq:fermionic-hamiltonian-modes}
\end{equation}
where the dispersion relation is the same as the one defined in eq.~(\ref{eq:dispersion-relation}), but the fermionic masses are now given by (\ref{eq:fermion-masses-GS}). 

\subsection{Dispersion relation and \texorpdfstring{$q\to1$}{q->1} limit}
\label{sec:NSNSdispersion}
The plane-wave dispersion relation~\eqref{eq:dispersion-relation} has the same form for bosons and fermions, even if for $\omega\neq\varphi$ the mass-spectrum of the excitations is not supersymmetric, $\mu_i^b\neq\mu_i^f$. For $0\leq q<1$ the dispersion we find is consistent with the one required by supersymmetry~\cite{Borsato:2012ud, Borsato:2014hja, Lloyd:2014bsa, Borsato:2015mma}, which takes the form\footnote{This can also be found from studying dionic giant magnons~\cite{David:2008yk,Hoare:2013lja}. Note that we have absorbed a factor of the tension into the definition of the momentum, {\it i.e.}\ eq.~(\ref{eq:exactdispersion}) leads to the usual dispersion relation by replacing $p\mapsto hp$.} 
\begin{equation}
\label{eq:exactdispersion}
\omega_p^\pm(\mu)=\sqrt{(q  p\pm \mu)^2+4(1-q^2)h^2\sin^2(p/2h)}\ ,
\end{equation}
where $h$ is the string tension. 
In the near-plane-wave limit the momentum is small relative to the tension, and we can approximate 
$4 h^2\sin^2(p/2h) \cong p^2$, where $p$ is the momentum from above, see eq.~(\ref{3.24}). 

The case $q=1$ is quite special since then the exact dispersion relation~\eqref{eq:exactdispersion} simplifies to the linear expression, which becomes in the plane-wave limit 
\begin{equation}
\omega_p^\pm(\mu)=\big| p\pm\mu\big| \approx \Big| \frac{2\pi\, n}{r} \pm \mu\Big|\ . 
\end{equation} 
The modulus signals that the world-sheet theory is chiral, \textit{i.e.}\ it distinguishes between left- and right-movers on the world-sheet. A left-moving mode is one which has $\partial\omega^\pm/\partial p>0$, which happens when $n>\mp \frac{\mu r}{2\pi}$. To illustrate this point, let us consider the contribution of a complex bosonic mode of mass~$\mu$ to the Hamiltonian. Suppressing the explicit $\mu$ dependence and setting $\nu \equiv \frac{\mu r}{2\pi}$, we get for the Hamiltonian
\begin{equation}
\begin{aligned}
&\sum_{n} \omega_n^+ \, a_n^{+ \dagger} a_n^{+} +\sum_{n} \omega_n^- \, a_n^{- \dagger} a_n^{-} \\
&\quad =\sum_{n>-\nu}\big(\tfrac{2\pi n}{r}+\mu\big) \, a_n^{+ \dagger} a_n^{+} 
+ \sum_{n<-\nu}\big(-\tfrac{2\pi n}{r}-\mu\big) \, a_n^{+ \dagger} a_n^{+}\\
&\quad \qquad\qquad\qquad\qquad\qquad\qquad\qquad
+\sum_{n>\nu}\big(\tfrac{2\pi n}{r}-\mu\big) \, a_n^{- \dagger} a_n^{-} 
+ \sum_{n<\nu}\big(-\tfrac{2\pi n}{r}+\mu\big) \, a_n^{- \dagger} a_n^{-}\\
&\quad =
\sum_{n>-\nu} \omega^+_n \, a_n^{+ \dagger} a_n^{+}
+ \sum_{\tilde{n}>\nu} \omega^-_{\tilde{n}} \, \tilde{a}_{\tilde{n}}^{- \dagger} \tilde{a}_{\tilde{n}}^{-}
+\sum_{n>\nu} \omega^-_n \, a_n^{- \dagger} a_n^{-} +
\sum_{\tilde{n}>-\nu} \omega^+_{\tilde{n}} \, \tilde{a}_{\tilde{n}}^{+ \dagger} \tilde{a}_{\tilde{n}}^{+}\,,
\end{aligned}
\end{equation}
where in the last line we have changed variables to $\tilde{n}=-n$ and set $\tilde{a}_{n}^\pm\equiv a_{-n}^\mp$.
Thus the complex boson of the GS description has split into two ``left'' (\textit{i.e.}\ holomorphic) real bosons $a^\pm$, and two ``right'' (\textit{i.e.}\ anti-holomorphic) real bosons $\tilde{a}^\pm$. Furthermore, their creation operators are shifted by $\pm \nu$. This shift was already observed by Maldacena and Ooguri~\cite{Maldacena:2000hw}, and, as we will see below, is closely related to the spectral flow. Indeed, loosely speaking the spectral flow parameter $w$ is the integer part of $\nu$.
Here we have tacitly assumed that $\nu=\frac{\mu r}{2\pi}\notin\mathbb{N}$; otherwise there are zero modes that require more care. 

\subsection{Ground-state energy}
It is worth noticing that the normal ordering constant
\begin{equation}
\label{eq:groundstateEn}
\mathcal{E}_0 = \sum_{i=1}^4 \displaystyle\sum_{n > 0} \left[\omega_n^+(\mu_i^b) + \omega_n^-(\mu_i^b) - \omega_n^+(\mu_i^f) - \omega_n^-(\mu_i^f)\right] + \sum_{i=1}^4 \left[ \mu_i^b - \mu_i^f\right]  
\end{equation}
does not vanish; this is due to the fact that the bosonic and fermionic masses, see eqs.~\eqref{eq:boson-masses-GS} and \eqref{eq:fermion-masses-GS}, are different. The resulting shift of the ground state energy is finite, as can be seen by observing that 
\begin{equation}
\sum_{i=1}^4 \big(\mu_i^b\big)^2 = \sum_{i=1}^4 \big(\mu_i^f \big)^2\,.
\end{equation}
For pure NS-NS flux, \textit{i.e.}\ $q = 1$, the summand of the first sum in eq.~\eqref{eq:groundstateEn} is independent of $n$, and $\mathcal{E}_0$ can be readily computed. In particular, if the masses $\mu_i$ satisfy $\mu_i< \frac{2\pi}{r}$, we find the simple expression
\begin{equation}
\mathcal{E}_0 = \sum_{i=1}^4 \left(\mu_i^b - \mu_i^f\right) = \cos \varphi \cos \omega + \sin \varphi \sin \omega-1=\cos(\varphi-\omega)-1\,. 
\label{eq:GS-normal-ordering-constant}
\end{equation}
This will play an important role below for the comparison with the WZW spectrum. 

\section{The WZW description}
\label{sec:the-wzw-spectrum}

String theory on $\AdSSSS$ with pure NS-NS flux can be described by a WZW model based on the $\mathcal{N} = 1$ affine algebras $\mathfrak{sl}(2)_k \oplus \mathfrak{sl}(2)_{k_1} \oplus \mathfrak{su}(2)_{k_2} \oplus \mathfrak{u}(1)$ \cite{Elitzur:1998mm,deBoer:1999gea,Eberhardt:2017fsi}; our conventions for the relevant algebras are described in appendix~\ref{app:superconformal-affine-algebras-and-spectral-flow-automorphism}. The levels $k$, $k_1$, $k_2$ are all positive integers, and are proportional to the radius squared of the corresponding space, {\it e.g.}\ $k\,\alpha' = \Rads^2$, and similarly for the $3$-spheres. In terms of our previous parametrization, see eq.~(\ref{2.4}), we therefore have 
\begin{equation}
\alpha = \dfrac{k}{k_1} = \cos^2 \varphi\ , \qquad 1-\alpha = \dfrac{k}{k_2} = \sin^2 \varphi\ . 
\end{equation}
The fermions can be decoupled from the Ka\v{c}-Moody currents as explained \textit{e.g.}\ in ref.~\cite{Ferreira:2017pgt}. The levels of the resulting $\mathfrak{sl}(2)$ and $\mathfrak{su}(2)$ affine algebras are then shifted by $+2$ and $-2$, respectively. We will denote the (decoupled) holomorphic currents  for $\mathfrak{sl}(2)_k$ by $K_{-n}^a$, and for the two $\mathfrak{su}(2)_{k_I}$ algebras by $J^{I,a}_{-n}$, where $I=1,2$ and $a \in \{+, -, 3 \}$. The corresponding fermions will be denoted by $\psi_{-r}^a$ and $\chi_{-r}^{I,a}$, respectively. We finally have an uncharged boson $\alpha_n$, together with an uncharged fermion $\eta_r$, coming from the $\mathfrak{u}(1)$ factor. The corresponding anti-holomorphic fields will be denoted by a tilde. 

The fermions are either half-integer moded (NS sector) or integer moded (R sector). In the unflowed sector --- there are also spectrally flowed sectors that will be explained in more detail below --- the representations of the affine algebras are conventional (Virasoro) highest weight representations generated from a ground state $\ket{j_0; j_{0,1}; j_{0,2}}$. Here $j_0$ labels a lowest weight state of $\mathfrak{sl}(2)$, while $j_{01}$ and $j_{02}$ are the spins with respect to the two $\mathfrak{su}(2)$ algebras, {\it i.e.}\ 
\begin{equation}
\begin{array}{lll}
K^a_{n} \ket{j_0; j_{0,1}; j_{0,2}} = 0\ , \qquad \qquad & J^{I,a}_{n} \ket{j_0; j_{0,1}; j_{0,2}} = 0\ , \qquad \qquad & n>0\ ,  \\
\psi_{r}^a \ket{j_0; j_{0,1}; j_{0,2}} = 0\ , \qquad \qquad & \chi_{r}^{I,a} \ket{j_0; j_{0,1}; j_{0,2}} = 0 \ , \qquad \qquad & r>0\ , \\
K^-_0 \ket{j_0; j_{0,1}; j_{0,2}} = 0\ , \qquad \qquad & J^{I,+}_0 \ket{j_0; j_{0,1}; j_{0,2}} = 0\ .  &
\end{array}
\end{equation}
The ground state representation is the same for left- and right-movers, and $j_0$ satisfies the Maldacena-Ooguri bound~\cite{Maldacena:2000hw} 
\begin{equation}
\frac{1}{2} < j_0 < \dfrac{k+1}{2}\ , 
\label{eq:maldacena-ooguri-bound}
\end{equation}
while unitarity requires for the $\mathfrak{su}(2)$ spins
\begin{equation}
0 \leq j_{0,I} \leq \dfrac{k_I -2}{2}\ . 
\label{eq:su(2)-unitarity-bound}
\end{equation}
The level matching condition is $N_{\rm{eff}} = \tilde N_{\rm{eff}}$, where 
\begin{equation}
N_{\rm{eff}} = \begin{cases} N - \tfrac{1}{2} & \rm{NS \ sector} \\ N & \rm{R \ sector,}  \end{cases}
\end{equation}
and $N$ is the total number operator. The spacetime energy and angular momenta are given by 
\begin{equation}
\Delta = j + \tilde{\jmath}\ , \qquad J_1 = j_1 + \tilde{\jmath}_1\ , \qquad J_2 = j_2 + \tilde{\jmath}_2\ ,
\label{eq:spacetime-charges} 
\end{equation}
where $j$ is the $\mathfrak{sl}(2)$ lowest weight of the relevant state, and similarly for the $\mathfrak{su}(2)$ spins. These quantum numbers differ in general from those of the ground states, and we define 
\begin{equation}
\begin{array}{lll}
j = j_0 + \delta j\ , \qquad & j_I = j_{0,I}  - \delta j_I \qquad  &\rm{NS \ sector} \\
j = j_0 + \delta j + s - \tfrac{1}{2}\ , \qquad  \qquad & j_I = j_{0,I}  - \delta j_I - s_I + \tfrac{1}{2} 
\qquad \qquad & \rm{R \ sector.}
\label{eq:WZW-spins}
\end{array}
\end{equation}
Thus $\delta j$ counts the number of $\mathfrak{sl}(2)$ modes with $a=+$ minus those with $a =-$, while $\delta j_I$ counts the number of $\mathfrak{su}(2)_I$ currents with $a = -$ minus those with $a = +$. In the R sector, $s, s_1, s_2 \in \{0, 1\}$ label the different ground states under the action of the fermionic zero-modes, see {\it e.g.}~\cite{Ferreira:2017pgt}. Physical states are annihilated by the positive super Virasoro modes, and obey the mass-shell condition
\begin{equation}
-\dfrac{j_0(j_0 - 1)}{k} + \dfrac{j_{0,1}(j_{0,1} + 1)}{k_1} + \dfrac{j_{0,2}(j_{0,2} + 1)}{k_2} + N_{\rm{eff}} = 0\ . 
\label{eq:unflowed-mass-shell-condition}
\end{equation}
Finally, the GSO projection requires that we have an odd number of fermionic excitations modes in the NS sector. (In the R sector, the GSO projection depends on whether we consider type IIA or type IIB string theory, but as we will see below, this does not affect the spacetime charges of the states.)

\subsection{Low-lying states} 
\label{subsec:low-lying-states}
In order to illustrate the construction of the spectrum, let us review briefly the structure of some of the low-lying states in the unflowed sector (including the BPS states).

\subsubsection{NS sector}

The GSO projection requires an odd number of fermions. Since they are half-integer moded, this is equivalent to $N_{\rm{eff}} \in \mathbb{N}$. The lowest lying states occur at level $N_{\rm{eff}} = 0$. We can solve the mass-shell condition \eqref{eq:unflowed-mass-shell-condition} for $j_0$ as a function of $j_{0,1}$ and $j_{0,2}$, obeying \eqref{eq:su(2)-unitarity-bound}. For each solution in the interval \eqref{eq:maldacena-ooguri-bound}, we then have a physical state. 

The physical states also have to be annihilated by the positive super Virasoro modes, and this leads to the following eight excitations 
\begin{equation}
\begin{aligned}
& \psi_{-\frac{1}{2}}^+ \ket{j_0; j_{0,1}; j_{0,2}}, & \qquad  &   \psi_{-\frac{1}{2}}^- \ket{j_0; j_{0,1}; j_{0,2}}, \\
& \chi_{-\frac{1}{2}}^{I,+} \ket{j_0; j_{0,1}; j_{0,2}}, & & \chi_{-\frac{1}{2}}^{I,-} \ket{j_0; j_{0,1}; j_{0,2}}, \\
 &  \eta_{-\frac{1}{2}} \ket{j_0; j_{0,1}; j_{0,2}}, & &  \hat{\eta}_{-\frac{1}{2}} \ket{j_0; j_{0,1}; j_{0,2}} \ , 
\end{aligned}
\end{equation}
where $\hat{\eta}$ is a certain linear combination of the $3$-modes $\psi^3$ and $\chi^{I,3}$ with $I=1,2$. 
The state $\psi_{-\frac{1}{2}}^- \ket{j_0; j_0-1; j_0-1}$ has $j = j_1 = j_2$ and is the only BPS state in the unflowed NS sector. 

\subsubsection{R sector}

In the R sector the different fermionic ground states are parameterized by the quantum numbers $s,s_1,s_2,s_3 \in \{0,1\}$, see \cite{Ferreira:2017pgt} for a careful discussion. Here $s$ arises from ${\rm AdS}_3$, while $s_{1,2}$ are associated with ${\rm S}^3_{1,2}$, and $s_3$ is the quantum number associated to the fermionic zero modes $\eta_0$ and $\hat{\eta}_0$. The value of $s_3$ does not have an effect on the spacetime charges (\ref{eq:spacetime-charges}), and hence we can always adjust $s_3\in\{0,1\}$ so as to satisfy the GSO projection. 
At level $N_{\rm{eff}} = 0$ we then find eight states, corresponding to the eight possible choices of $(s,s_1,s_2)$. We can label them as 
\begin{equation}
\ket{j_0, j_{0,1}, j_{0,2}}_{(s,s_1,s_2)}\ . 
\end{equation}  
The state $\ket{j_0; j_0-1; j_0-1}_{(0,0,0)}$ has $j = j_1 = j_2$ and is the only BPS state arising in the unflowed R sector \cite{Eberhardt:2017fsi}. 

\subsection{Spectral flow}
\label{subsec:spectral-flow}

Because of the Maldacena-Ooguri bound (\ref{eq:maldacena-ooguri-bound}), the spacetime spectrum arising from the unflowed sector leads to states with a finite maximal energy. In order to overcome this unphysical feature, it was argued in \cite{Maldacena:2000hw} that the spectrum must also contain the spectrally flowed images of these representations.\footnote{The significance of spectral flow was first noticed in \cite{Henningson:1991jc} based on modular invariance arguments.} Spectral flow is an outer automorphism of the affine algebra $\mathfrak{sl}(2)_k$, see appendix~\ref{app:superconformal-affine-algebras-and-spectral-flow-automorphism} for the explicit description, and hence relates in general inequivalent representations to one another. For $\mathfrak{su}(2)_k$ one can also define a spectral flow automorphism, but this does not lead to new representations. However, it is sometimes convenient to flow also in the $\mathfrak{su}(2)_{k_I}$ factors, as we shall do below.

Since spectral flow shifts the eigenvalues of the Cartan generators, see eq.~\eqref{eq:spectrally-flowed-generators}, we have now instead of  \eqref{eq:WZW-spins}
\begin{equation}
\begin{array}{lll}
j = \frac{k w}{2} + j_0 + \delta j \ , \qquad \qquad & j_I = \frac{k_I w_I}{2} + j_{0,I}  - \delta j_I\qquad \qquad & \rm{NS \ sector} \\
j = \frac{k w}{2} + j_0 + \delta j + s - \tfrac{1}{2}\ , \qquad \qquad & j_I = \frac{k_I w_I}{2} + j_{0,I}  - \delta j_I - s_I + \tfrac{1}{2} \qquad \qquad & \rm{R \ sector.}
\label{eq:flowed-WZW-spins}
\end{array}
\end{equation}
The spectral flow also affects the eigenvalue of $L_0$, see eq.~(\ref{eq:spectrally-flowed-generators}), and hence the mass-shell condition, eq.~(\ref{eq:unflowed-mass-shell-condition}), becomes in the spectrally flowed NS sector
\begin{multline}
-\frac{j_0 (j_0 - 1)}{k} - w (j_0 + \delta j) - \frac{k}{4}w^2 + \frac{j_{0,1}(j_{0,1} + 1)}{k_1} + w_1(j_{0,1} - \delta j_1) + \frac{k_1}{4} (w_1)^2 \\ 
+ \frac{j_{0,2}(j_{0,2} + 1)}{k_2} + w_2(j_{0,2} - \delta j_2) + \frac{k_2}{4} (w_2)^2 + N_{\rm{eff}} = 0\ , 
\label{eq:flowed-mass-shell-condition-NS}
\end{multline}
and in the spectrally flowed R sector 
\begin{multline}
-\frac{j_0 (j_0 - 1)}{k} - w (j_0 + \delta j -\tfrac{1}{2}) - \frac{k}{4}w^2 + \frac{j_{0,1}(j_{0,1} + 1)}{k_1} + w_1(j_{0,1} - \delta j_1 + \tfrac{1}{2}) + \frac{k_1}{4} (w_1)^2 \\ 
+ \frac{j_{0,2}(j_{0,2} + 1)}{k_2} + w_2(j_{0,2} - \delta j_2 + \tfrac{1}{2}) + \frac{k_2}{4} (w_2)^2 + N_{\rm{eff}} = 0 \ . 
\label{eq:flowed-mass-shell-condition-R}
\end{multline}
Furthermore, the level matching condition now becomes
\begin{equation}
N_{\rm{eff}} - w \delta j - w_1 \delta j_1 - w_2 \delta j_2 = \tilde{N}_{\rm{eff}} - w \delta \tilde \jmath - w_1 \delta \tilde \jmath_1 - w_2 \delta \tilde \jmath_2\ , 
\end{equation}
where $N_{\rm{eff}}$ and $\tilde{N}_{\rm{eff}}$ are the number operators before spectral flow. 
In the R sector the GSO projection can still be satisfied using (if necessary) the action of the fermionic zero modes $\eta_0$ and $\hat{\eta}_0$, while in the NS sector it reads \cite{Giribet:2007wp, Ferreira:2017pgt}
\begin{equation}
N_{\rm{eff}} + \frac{w + w_1 + w_2}{2} \in \mathbb{N}\ . 
\end{equation}
The construction of the BPS states in the spectrally flowed sectors is quite involved for $\AdSSSS$, and we refer the reader to \cite{Eberhardt:2017fsi, Eberhardt:2017pty} for a detailed explanation. 

\section{Plane-wave limit of the WZW spectrum}
\label{sec:plane-wave-limit}

In this subsection we study the plane-wave limit of the WZW model for $\AdSSSS$, generalising the analysis of  \cite{Son:2003zv} for AdS$_3 \times$S$^3 \times$T$^4$. We will consider the limit in which both the levels $k$, $k_1$, $k_2$ are large (so that we are in the supergravity regime), as well as the charges $\Delta$, $J_1$, $J_2$ (as dictated by the Penrose limit, see Section~\ref{subsec:the-penrose-limit}). Note that it follows from \eqref{eq:null-condition-solution} that all these charges scale then as $k\sim R^2$.

Since we are interested in states that are obtained from the ground states by a finite number of creation operators, the actual eigenvalues of the states differ only by a finite (and hence small) amount from those of the ground states. Since the ground states have the same spin for left- and right-movers --- this is true both for the $\mathfrak{sl}(2,\mathbb{R})$ lowest weight, as well as for the two $\mathfrak{su}(2)$ spins --- all spins scale the same way for left- and right-movers,  {\it i.e.}\ $j \sim \tilde \jmath$ and $j_I \sim \tilde \jmath_I$; this is 
also required by the form of the geodesic~\eqref{eq:geodesic}. 

In the Penrose limit, {\it i.e.}\ up to subleading corrections in $1/k$,  it follows from eq.~\eqref{eq:p_-propto-delta} that 
\begin{equation}
P_- = r =2\pi\alpha'\frac{\Delta}{R^2}= \dfrac{4\pi j}{k}\ . 
\label{eq:p_-=...WZW}
\end{equation}
Note that the Maldacena-Ooguri bound, eq.~(\ref{eq:maldacena-ooguri-bound}), implies that $P_{-}\leq 2\pi$, but as mentioned above, the spectrum also contains spectrally flowed sectors that effectively shift up the spin. Indeed, in the sector with spectral flow $w$ we have 
\begin{equation}
2 \pi w \lesssim r \lesssim 2 \pi(w+1)\ .
\label{eq:p_-values-flowed-sector}
\end{equation}
Note that the different `sectors' are also visible from the viewpoint of the plane-wave dispersion relation \eqref{eq:dispersion-relation}: at $q=1$, {\it i.e.}\ for pure NS-NS flux, it can vanish when $r=P_-=0$ mod~$2\pi$. 
As $P_-$ crosses such a value, it is necessary to adjust the definition of positive- and negative-energy modes in the light-cone Hamiltonian. As a consequence, the spectrum is composed of different sectors depending on ``how fast'' the geodesic is spinning, \textit{cf.}\ section~\ref{sec:NSNSdispersion}. 

It is therefore convenient to discuss the WZW spectrum separately for the case where the geodesic is \emph{slowly spinning}, so that $w=0$ and we can use the formulae for the ``unflowed'' spectrum; and the case where the geodesic is \emph{rapidly spinning}, where we will need the more general formulae of section~\ref{subsec:spectral-flow}.

\subsection{Slowly spinning geodesics (unflowed sector)}

In the following we shall analyse the spacetime spectrum of string theory for large level~$k$, as well as for large spins~$j,j_I$.\footnote{In the unflowed sector corresponding to slowly spinning strings, we have however always $\frac{j}{k}\leq \frac{1}{2}$.} Since we are interested in the Penrose limit of a specific null geodesic, we have in addition, see eq.~\eqref{eq:null-condition-solution},
\begin{equation}
j_1 \sim j\, \dfrac{\cos \omega}{\cos \varphi}\,, \qquad j_2 \sim j\, \dfrac{\sin \omega}{\sin \varphi}\ , 
\label{eq:chiral-null-geodesic-solution}
\end{equation}
so that
\begin{equation}
j_1^2\,\cos^2 \varphi  +  j_2^2\,\sin^2 \varphi \sim j^2\ , \qquad 
j_1\,\cos \varphi \cos \omega  + j_2\,\sin \varphi \sin \omega \sim j\ . 
\label{eq:j-relations-penrose-limit}
\end{equation}
This will further simplify our expressions.

\subsubsection{R sector}

Let us begin by studying the R sector.  We start from the mass-shell formula \eqref{eq:unflowed-mass-shell-condition} 
\begin{equation}
j_0 = \frac{1}{2} + \sqrt{k N_{\rm{eff}} +  \left( j_{0,1} + \tfrac{1}{2} \right)^2 \, \cos^2 \varphi
+   \left( j_{0,2} + \tfrac{1}{2} \right)^2\, \sin^2 \varphi} \ , 
\end{equation}
which we rewrite, using \eqref{eq:WZW-spins}, as 
\begin{multline}
j = \delta j + s + \Big[  j_1^2\, \cos^2 \varphi  +  j_2^2\, \sin^2 \varphi  +
k N_{\rm{eff}} + 2  j_1(\delta j_1 + s_1) \, \cos^2 \varphi  +  2  j_2(\delta j_2 + s_2)\, \sin^2 \varphi  \\
 + (\delta j_1 + s_1)^2 + (\delta j_2 + s_2)^2 \Big]^{\frac{1}{2}}. 
\label{eq:mass-shell-formula-solution-true-spins-unflowed-R}
\end{multline}
The holomorphic contribution to the light-cone Hamiltonian $H=-P_+$ is then, see eq.~(\ref{P+}),
\begin{equation}
H_L = j -  j_1 \cos \varphi \cos \omega  - j_2 \sin \varphi \sin \omega  \ . 
\end{equation} 
For large $k$ and large charges $j, j_1, j_2$, the leading term of the square root in (\ref{eq:mass-shell-formula-solution-true-spins-unflowed-R}) is 
\begin{equation}
j_1^2\, \cos^2 \varphi  +  j_2^2\, \sin^2 \varphi \sim j^2 \ . 
\end{equation}
Taking this combination out of the square root and expanding the remaining terms to leading order, we then obtain, 
\begin{equation}
H_L =  \delta j + s + \frac{2 \pi N_{\rm{eff}}}{r}  + \left(\delta j_1 + s_1 \right)\, \cos \varphi \cos \omega  +  \left(\delta j_2 + s_2  \right)\, \sin \varphi \sin \omega \ , 
\label{eq:unflowed-R-hamiltonian}
\end{equation}
where we have written $\frac{ k \, N_{\rm{eff}}}{2 j } = \frac{2 \pi N_{\rm{eff}}}{r}$, see eq.~(\ref{eq:p_-=...WZW}).

Essentially the same formula also holds for the anti-holomorphic contribution $H_R$. Note that since 
$\delta j$ and $\delta j_I$ can take both positive and negative values, one may worry that the Hamiltonian \eqref{eq:unflowed-R-hamiltonian} is not bounded from below. However this is not the case, as can be checked directly; in fact, even before taking the Penrose limit, the non-linear WZW energy is also bounded from below \cite{Lorenz}, and this therefore also remains true in the Penrose limit.

\subsubsection{NS sector}

The analysis in the unflowed NS sector is essentially the same. Again, we rewrite the mass-shell formula \eqref{eq:unflowed-mass-shell-condition} in terms of the ``true spins'' $j, j_I$ (rather than the spins $j_0, j_{0,I}$ of the ground state), using eq.~\eqref{eq:WZW-spins}
\begin{equation}
j = \delta j + \tfrac{1}{2} + \Big[  ( j_1 + \delta j_1 + \tfrac{1}{2} )^2\, \cos^2 \varphi  +   ( j_2 + \delta j_2 + \tfrac{1}{2})^2 \, \sin^2 \varphi  + k N_{\rm{eff}}  \Big]^{\frac{1}{2}}\ .
\label{eq:mass-shell-formula-solution-true-spins-unflowed-NS}
\end{equation}
In the Penrose limit, the holomorphic part of the light-cone Hamiltonian $H=-P_+$ then becomes
\begin{equation}
H_L = \dfrac{2 \pi N_{\rm{eff}}}{r} + \delta j + \tfrac{1}{2} + \left(\delta j_1 + \tfrac{1}{2} \right) \cos \varphi \cos \omega  + \left(\delta j_2 + \tfrac{1}{2}  \right) \sin \varphi \sin \omega \ ,  
\label{eq:unflowed-NS-hamiltonian}
\end{equation}
and similarly for $H_R$. We note that, due to the GSO projection, at least one fermion has to act on the ground state. 

\subsection{Rapidly-spinning geodesics (flowed sector)}
\label{sec:flowed-sector}

Fast spinning solutions are characterised by the condition that $j>\frac{k}{2}$, so that they come from sectors with spectral flow $w>0$, see eq.~(\ref{eq:p_-values-flowed-sector}). It is convenient to spectrally flow also in the two $\mathfrak{su}(2)$ sectors, where the relevant spectral flows are determined by 
\begin{equation}
\begin{aligned}
2 \pi w_1 \lesssim&\ \cos \varphi \cos \omega &\!\!\!r\ &\lesssim 2 \pi (w_1 + 1)\ , \\
2 \pi w_2 \lesssim&\ \sin \varphi \sin \omega &\!\!\!r\ &\lesssim 2 \pi (w_2 + 1)\ .  
\end{aligned}
\end{equation}
In particular, this guarantees that $\delta j_I$ is of order one, and hence we can proceed very similarly as in the unflowed sector. We shall again first deal with the R sector.

\subsubsection{R sector}

Using eq.~\eqref{eq:flowed-WZW-spins}, the mass-shell condition \eqref{eq:flowed-mass-shell-condition-R} in the spectrally flowed R sector can be rewritten in terms of the ``true spins'' $j, j_I$ as
\begin{multline}
k(N_{\rm{eff}} - w \delta j - w_1 \delta j_1 - w_2 \delta j_2) - (j - \delta j - s)^2 \\
+  (j_1 + \delta j_1 + s_1)^2\, \cos^2 \varphi +  (j_2 + \delta j_2 + s_2)^2\, \sin^2 \varphi =  0\ ,
\end{multline} 
which can be solved for $j$ to give 
\begin{multline}
j = \delta j + s + \Big[  (j_1 + \delta j_1 + s_1)^2\, \cos^2 \varphi + (j_2 + \delta j_2 + s_2)^2\, \sin^2 \varphi  \\
+ k(N_{\rm{eff}} - w \delta j - w_1 \delta j_1 - w_2 \delta j_2) \Big]^{\frac{1}{2}} \ . 
\end{multline}
The leading terms in the large $k$ and large charge limit are the quadratic charge terms in the first line of the square root. Proceeding as in the unflowed sector we therefore find in the Penrose limit 
\begin{multline}\label{eq:flowed-R-hamiltonian}
H_L = \dfrac{2 \pi (N_{\rm{eff}} - w \delta j - w_1 \delta j_1 - w_2 \delta j_2)}{r} + \delta j + s \\
+ \left(\delta j_1 + s_1 \right)\, \cos \varphi \cos \omega  + \left(\delta j_2 + s_2  \right)\, \sin \varphi \sin \omega \ .  
\end{multline}
Again one can convince oneself that the Hamiltonian is bounded from below, as must be the case.

\subsubsection{NS sector}

In the NS flowed sector the analysis works completely similarly. Using now \eqref{eq:flowed-WZW-spins} we find from the spectrally flowed mass-shell condition in the NS sector 
\begin{multline}
j = \delta j + \frac{1}{2} + \Big[ (j_1 + \delta j_1 + \tfrac{1}{2})^2 \, \cos^2 \varphi 
+ (j_2 + \delta j_2 + \tfrac{1}{2})^2 \,  \sin^2 \varphi   \\ 
+ k \Big(N_{\rm{eff}} - \frac{w + w_1 + w_2}{2} - w \delta j - w_1 \delta j_1 - w_2 \delta j_2 \Big) \Big]^{\frac{1}{2}} \ . 
\end{multline}
The holomorphic contribution to the Hamiltonian in the Penrose limit then becomes
\begin{multline}\label{eq:flowed-NS-hamiltonian}
H_L = 2 \pi \dfrac{N_{\rm{eff}} - \tfrac{1}{2}(w + w_1 + w_2) - w \delta j -  w_1 \delta j_1 -  w_2 \delta j_2}{r} + \delta j + \tfrac{1}{2} \\
+ (\delta j_1 + \tfrac{1}{2})\, \cos \varphi \cos \omega  + ( \delta j_2 + \tfrac{1}{2})\, \sin \varphi \sin \omega \ . 
\end{multline}

\section{Comparison of the spectra at \texorpdfstring{$q=1$}{q=1}}
\label{sec:comparison}

Now we have everything at our disposal to compare the Hamiltonian in the Penrose limit of the WZW spectrum that we have just derived, see eqs.~(\ref{eq:unflowed-R-hamiltonian}), 
(\ref{eq:unflowed-NS-hamiltonian}), (\ref{eq:flowed-R-hamiltonian}), and (\ref{eq:flowed-NS-hamiltonian}), with the 
one that was obtained from the GS formulation at $q=1$ in section~\ref{sec:the-green-schwarz-action}, see
eqs.~(\ref{eq:bosonic-hamiltonian-modes}) and (\ref{eq:fermionic-hamiltonian-modes}). As we shall see, this is quite straightforward for the bosonic excitations, but requires a little more work for states that involve fermions. 

We shall first consider the states at the lowest level, $N_{\rm{eff}} = 0$, before discussing systematically the general case. 
The GS Hamiltonian $H$ of eqs. \eqref{eq:bosonic-hamiltonian-modes} and \eqref{eq:fermionic-hamiltonian-modes}, splits naturally into a holomorphic and an anti-holomorphic part
\begin{equation}
H = H_L + H_R\ , 
\end{equation}  
and it is hence sufficient to compare (and match) the spectrum for the holomorphic (and the anti-holomorphic) part separately. For $r<2\pi$ (corresponding to the unflowed sector in the WZW model) the holomorphic part of the GS Hamiltonian at $q=1$  can be written as 
\begin{equation}
\label{eq:left-hamilt-gs}
\begin{aligned}
H_L = & \displaystyle\sum_{i=1}^4 \displaystyle\sum_{n >0} \Big( |n + \mu_i^f| \, b_n^{i,+ \dagger} b_n^{i,+} + |n - \mu_i^f| \, b_n^{i, - \dagger} b_n^{i,-}\Big) \\
&  + \displaystyle\sum_{i=1}^4 \displaystyle\sum_{n >0} \Big( |n + \mu_i^b| \, a_n^{i,+ \dagger} a_n^{i,+} + |n-\mu_i^b| \, a_n^{i,- \dagger} a_n^{i,-} \Big) \\
& +  \displaystyle\sum_{i=1}^4 \mu_i^f b_0^{i,+ \dagger} b_0^{i,+} + \displaystyle\sum_{i=1}^4 \mu_i^b a_0^{i,+ \dagger} a_0^{i,+}  + \frac{-1 + \cos \varphi \cos \omega + \sin \varphi \sin \omega}{2} \ .
\end{aligned}
\end{equation}
The anti-holomorphic part has essentially the same form, except that the sums run instead over $n<0$, and that the zero mode part only involves the $a_0^{i,-}$ and  $b_0^{i,-}$ generators (instead of $a_0^{i,+}$ and  $b_0^{i,+}$).

The bosonic oscillators can now be identified straightforwardly, and for the holomorphic sector we have written out the details in Table~\ref{tab:bosonic-modes-identifications}; a similar identification also applies to the anti-holomorphic generators.

\renewcommand{\arraystretch}{1.5}
\begin{table}[h]
\centering
\begin{tabular}{|c|c|c|}
\hline
WZW mode & GS mode & Plane-wave energy shift \\
\hline\hline
$K_{-n}^\pm$ & $a_n^{1, \pm \dagger}$ & $\frac{2\pi n}{r} \pm 1$ \\
\hline
$J_{-n}^{1,\mp}$ & $a_n^{2, \pm \dagger}$ & $\frac{2\pi n}{r} \pm \cos \varphi \cos \omega$ \\
\hline
$J_{-n}^{2,\mp}$ & $a_n^{3, \pm \dagger}$ & $\frac{2\pi n}{r} \pm \sin \varphi \sin \omega$ \\
\hline
$\alpha_{-n}^{1}, \alpha_{-n}^{2}$ & $a_n^{4, \pm \dagger}$ & $\frac{2\pi n}{r}$ \\
\hline\hline
$K_0^+$ & $a_0^{1, + \dagger}$ & $1$ \\
\hline
$J_0^{1,-}$ & $a_0^{2, + \dagger}$ & $\cos \varphi \cos \omega$ \\
\hline
$J_0^{2,-}$ & $a_0^{3, + \dagger}$ & $ \sin \varphi \sin \omega$ \\
\hline
$\alpha_0^{1}$ & $a_0^{4, + \dagger}$ & $0$ \\
\hline
\end{tabular}
\caption{Identification of the bosonic modes in the holomorphic sector.
The modes of the left-moving currents with $n>0$ are related to GS oscillators with $n>0$. 
}
\label{tab:bosonic-modes-identifications}
\end{table}
\renewcommand{\arraystretch}{1}

\subsection{The identification of the fermions}

The identification for the fermions is a bit more involved, reflecting the usual difficulty in relating the NS-R formulation (as in the WZW model) to the GS formulation. In order to understand how things work out, it is instructive to describe first the lowest lying states systematically. They arise from the unflowed sector of the WZW model at level $N_{\rm{eff}} = 0$.

\subsubsection{Level \texorpdfstring{$N_{\text{eff}} = 0$}{Neff=0}}
\label{subsec:level-0-matching}

WZW states at level $N_{\rm{eff}} = 0$ correspond to GS states built acting only with fermionic zero-modes. More precisely the NS sector states will be identified with the GS states involving an even number of GS zero modes, while the R sector states will come from an odd number of GS zero-modes.

More specifically, in the R sector at level $N_{\rm{eff}} = 0$ the WZW Hamiltonian in eq. \eqref{eq:unflowed-R-hamiltonian} simplifies to 
\begin{equation}
H_L = s + s_1 \cos \varphi \cos \omega + s_2 \sin \varphi \sin \omega\ . 
\end{equation}  
Here $s$, $s_1$ and $s_2$ take the values $\{0,1\}$, and the resulting eight states have been tabulated in 
Table~\ref{tab:spectrum-comparison-R-sector}, together with the GS states to which they correspond. 
 
\renewcommand{\arraystretch}{1.7}
\begin{table}[h]
\centering
\begin{tabular}{|c|c|c|c|}
\hline
WZW state & GS state & $H_L$ & $Z$\\
\hline 
$(s, s_1, s_2)= (0,0,0)$ & $b_0^{4,+ \dagger}\ket{0}$ & 0 & $u^{-\frac{1}{2}}y^{-\frac{1}{2}}z^{-\frac{1}{2}}$ \\
\hline
$(s, s_1, s_2)= (0,0,1)$ & $b_0^{3,+ \dagger}\ket{0}$ & $\sin \varphi \sin \omega$ & $u^{-\frac{1}{2}}y^{-\frac{1}{2}}z^{\frac{1}{2}}$ \\
\hline 
$(s, s_1, s_2)= (0,1,0)$ & $b_0^{2,+ \dagger}\ket{0}$ & $\cos \varphi \cos \omega$ & $u^{-\frac{1}{2}}y^{\frac{1}{2}}z^{-\frac{1}{2}}$  \\
\hline 
$(s, s_1, s_2)= (0,1,1)$ & $b_0^{1,+ \dagger}\ket{0}$ & $\sin \varphi \sin \omega + \cos \varphi \cos \omega$ & $u^{-\frac{1}{2}}y^{\frac{1}{2}}z^{\frac{1}{2}}$ \\
\hline 
$(s, s_1, s_2)= (1,0,0)$ & $b_0^{2,+ \dagger}b_0^{3,+ \dagger}b_0^{4,+ \dagger}\ket{0}$ & 1 & $u^{\frac{1}{2}}y^{-\frac{1}{2}}z^{-\frac{1}{2}}$ \\
\hline 
$(s, s_1, s_2)= (1,0,1)$ & $b_0^{1,+ \dagger}b_0^{3,+ \dagger}b_0^{4,+ \dagger}\ket{0}$ & $1 + \sin \varphi \sin \omega$ & $u^{\frac{1}{2}}y^{-\frac{1}{2}}z^{\frac{1}{2}}$ \\
\hline
$(s, s_1, s_2)= (1,1,0)$ & $b_0^{1,+ \dagger}b_0^{2,+ \dagger}b_0^{4,+ \dagger}\ket{0}$ & $1 + \cos \varphi \cos \omega$ & $u^{\frac{1}{2}}y^{\frac{1}{2}}z^{-\frac{1}{2}}$ \\
\hline 
$(s, s_1, s_2)= (1,1,1)$ & $b_0^{1,+ \dagger}b_0^{2,+ \dagger}b_0^{3,+ \dagger}\ket{0}$ & $1 + \sin \varphi \sin \omega + \cos \varphi \cos \omega$ & $u^{\frac{1}{2}}y^{\frac{1}{2}}z^{\frac{1}{2}}$  \\
\hline
\end{tabular}
\caption{Low-lying states in the R sector. In the left column we describe the states in terms of the fermionic zero-modes in the WZW language. In the second column, we describe them in term of GS fermionic oscillators; note that R sector states correspond to an odd number of GS fermions. In the remaining two columns we write the contribution to the (holomorphic) Hamiltonian and to the partition function that is described below.}
\label{tab:spectrum-comparison-R-sector}
\end{table}
\renewcommand{\arraystretch}{1}

\medskip

\noindent In the NS sector, the WZW Hamiltonian of eq. \eqref{eq:unflowed-NS-hamiltonian} simplifies to 
\begin{equation}
H_L = \delta j + \tfrac{1}{2} + (\delta j_1 + \tfrac{1}{2}) \cos \varphi \cos \omega + (\delta j_2 + \tfrac{1}{2}) \sin \varphi \sin \omega\ . 
\end{equation}
There are again $8$ states, and as explained in Table~\ref{tab:spectrum-comparison-NS-sector}, they can be identified with GS states built by applying an even number of fermionic zero-modes. 

\renewcommand{\arraystretch}{2}
\begin{table}[h]
\centering
\begin{tabular}{|c|c|c|c|}
\hline
WZW state & GS state & $H_L$ & $Z$\\
\hline 
$\psi^-_{-\frac{1}{2}} \ket{j_0; j_{0,1}; j_{0,2}}$ & $\ket{0}$ & $\frac{-1 + \cos \varphi \cos \omega + \sin \varphi \sin \omega}{2}$ & $u^{-1}$  \\
\hline 
$\chi^{1,+}_{-\frac{1}{2}} \ket{j_0; j_{0,1}; j_{0,2}}$ & $b_0^{3,+ \dagger}b_0^{4,+ \dagger}\ket{0}$ & $\frac{1 - \cos \varphi \cos \omega + \sin \varphi \sin \omega}{2}$ & $y^{-1}$ \\ 
\hline
$\chi^{2,+}_{-\frac{1}{2}} \ket{j_0; j_{0,1}; j_{0,2}}$ & $b_0^{2,+ \dagger}b_0^{4,+ \dagger}\ket{0}$ & $\frac{1 + \cos \varphi \cos \omega - \sin \varphi \sin \omega}{2}$ & $z^{-1}$ \\ 
\hline
$\langle\eta_{-\frac{1}{2}}, \,  \hat \eta_{-\frac{1}{2}}\rangle  \ket{j_0; j_{0,1}; j_{0,2}}$ & $\langle b_0^{2,+ \dagger}b_0^{3,+ \dagger},\, b_0^{1,+ \dagger}b_0^{4,+ \dagger}\rangle\ket{0}$ & $\frac{1 + \cos \varphi \cos \omega + \sin \varphi \sin \omega}{2}$ & 2 \\ 
\hline
$\chi^{2,-}_{-\frac{1}{2}} \ket{j_0; j_{0,1}; j_{0,2}}$ & $b_0^{1,+ \dagger}b_0^{3,+ \dagger}\ket{0}$ & $\frac{1 + \cos \varphi \cos \omega +3 \sin \varphi \sin \omega}{2}$ & $z$ \\ 
\hline
$\chi^{1,-}_{-\frac{1}{2}} \ket{j_0; j_{0,1}; j_{0,2}}$ & $b_0^{1,+ \dagger}b_0^{2,+ \dagger}\ket{0}$ & $\frac{1 + 3 \cos \varphi \cos \omega + \sin \varphi \sin \omega}{2}$ & $y$ \\ 
\hline
$\psi^+_{-\frac{1}{2}} \ket{j_0; j_{0,1}; j_{0,2}}$ & $b_0^{1,+ \dagger}b_0^{2,+ \dagger} b_0^{3,+ \dagger}b_0^{4,+ \dagger}\ket{0}$ & $\frac{3 + \cos \varphi \cos \omega + \sin \varphi \sin \omega}{2}$ & $u$\\ 
\hline
\end{tabular}
\caption{Low-lying states in the NS sector. In the left column we describe the states in terms of the fermionic modes in the WZW language. In the second column, we describe them in terms of the GS fermionic zero-modes; note that NS sector states correspond to an even number of GS fermions. In the remaining two columns we write the contribution to the (holomorphic) Hamiltonian and to the partition function described below.}
\label{tab:spectrum-comparison-NS-sector}
\end{table}
\renewcommand{\arraystretch}{1}

\subsection{The general case}
\label{subsec:higher-level-match}

At higher level, an explicit identification as above becomes cumbersome, but we can demonstrate that the spectra match by considering the generating function (with suitable chemical potentials). More specifically, let us  define in the WZW model 
\begin{equation}
Z(q,u,y_1,y_2) = \sum_{N,j,j_1,j_2} d(N,j,j_1,j_2) \, q^N\, u^{(j-j_0)}\, y_1^{(j_1-j_{0,1})} \, y_2^{(j_2-j_{0,2})} \ , 
\end{equation}
where $d(N,j,j_1,j_2)$ is the multiplicity of states with $N=N_{\rm{eff}}$ and spins $j$, $j_1$ and $j_2$; as we are interested in the plane-wave limit, it makes sense to subtract out the spins of the ground states, {\it i.e.}\ 
$j_0$ and $j_{0,I}$. In the following we shall only keep track of the states that are generated by the fermions; the contribution of the bosons matches directly, see Table~\ref{tab:bosonic-modes-identifications}.

\noindent In the NS sector we then find 
\begin{align}
Z_{{\rm NS}}  =\, \frac{1}{2} \, \Bigl[ & \prod_{n=1}^{\infty} (1 +  u\, q^{n-\frac{1}{2}}) \, (1 +  u^{-1}\, q^{n-\frac{1}{2}})\,(1 + q^{n-\frac{1}{2}})^2 \, \times \nonumber \\
& \qquad \times (1 +  y_1\, q^{n-\frac{1}{2}}) \, (1 +  y_1^{-1}\, q^{n-\frac{1}{2}}) \, (1 +  y_2\, q^{n-\frac{1}{2}}) \, (1 +  y_2^{-1}\, q^{n-\frac{1}{2}})  \label{eq:z-ns} \\
& - \prod_{n=1}^{\infty} (1 -  u\, q^{n-\frac{1}{2}}) \, (1 -  u^{-1}\, q^{n-\frac{1}{2}})\,(1 - q^{n-\frac{1}{2}})^2\, \times \nonumber \\
& \qquad \times (1 -  y_1\, q^{n-\frac{1}{2}}) \, (1 -  y_1^{-1}\, q^{n-\frac{1}{2}}) \, (1 -  y_2\, q^{n-\frac{1}{2}}) \, (1 -  y_2^{-1}\, q^{n-\frac{1}{2}}) \Bigr] \ , \nonumber
\end{align}
where the difference of the two terms implements the GSO projection, while the R sector contribution is 
\begin{multline}
Z_{{\rm R}}  =\, q^{\frac{1}{2}} (u^{-\frac{1}{2}} + u^{\frac{1}{2}})(y_1^{-\frac{1}{2}}+y_1^{\frac{1}{2}})(y_2^{-\frac{1}{2}}+y_2^{\frac{1}{2}}) \, \prod_{n=1}^{\infty} (1 +  u\, q^n) \, (1 +  u^{-1}\, q^n)\,(1 + q^n)^2\, \times \\
 \qquad \times (1 +  y_1\, q^n) \, (1 +  y_1^{-1}\, q^n)  (1 +  y_2\, q^n) \, (1 +  y_2^{-1}\, q^n)\,\ .
\end{multline}
It follows from the abstruse identity, see {\it e.g.}\ \cite[Chapter 21]{whittaker1996course},  that 
\begin{equation}\label{abstruse}
Z_{{\rm NS}}  + Z_{{\rm R}}  = Z_{{\rm GS}}  \ , 
\end{equation}
where 
\begin{equation}
Z_{{\rm GS}}  =  q^{\frac{1}{2}} \prod_{i=1}^4 (z_i^{\frac{1}{2}} + z_i^{-\frac{1}{2}}) \, 
\prod_{n=1}^{\infty} (1 +  z_i \, q^n) \, (1 +  z_i^{-1} \, q^n) \ , 
\end{equation}
and the $z_i$ are defined as 
\begin{equation}
\begin{aligned}
&z_1 = u^{\frac{1}{2}} y^{\frac{1}{2}} z^{\frac{1}{2}} \ , & \qquad  & z_2 = u^{\frac{1}{2}} y^{\frac{1}{2}} z^{-\frac{1}{2}}  \ , \\
&z_3 = u^{\frac{1}{2}} y^{-\frac{1}{2}} z^{\frac{1}{2}}\ , & \qquad  & z_4 =u^{\frac{1}{2}} y^{-\frac{1}{2}} z^{-\frac{1}{2}} \ . 
\end{aligned}
\end{equation} 

In order to compare the WZW spectrum with that obtained from the  GS formalism we now expand the WZW spectrum according to its $H_L$ eigenvalues, {\it i.e.}\ we consider 
\begin{equation}\label{zidef}
Z\Bigl(q=e^{2\pi i \tau}, \, u= e^{2 \pi i \mu_1^b \tau}, \, y_1 = e^{2 \pi i \mu_2^b \tau}, \, y_2 = e^{2 \pi i \mu_3^b \tau} \Bigr) = \sum_{H} d(H) e^{2\pi i \tau H} \ , 
\end{equation}
where $d(H)$ is now the multiplicity of states with $H_L=H$; here we have used the form of the light-cone Hamiltonian in the unflowed sector, see eq.~(\ref{eq:unflowed-R-hamiltonian}) and (\ref{eq:unflowed-NS-hamiltonian}). It now follows from the specific form of the bosonic and fermionic mass eigenvalues, see eqs.~(\ref{eq:boson-masses-GS}) and (\ref{eq:fermion-masses-GS}), that with these chemical potentials, the $z_i$ parameters defined in (\ref{zidef}) equal 
\begin{equation}
z_i = e^{2\pi i \mu_i^f \, \tau} \ .
\end{equation}
Thus the right-hand-side of the abstruse identity (\ref{abstruse}) gives directly the GS spectrum. This proves the matching of the spectrum in the unflowed sector. Since the light-cone Hamiltonian of the flowed sectors have essentially the same form as in the unflowed sector, {\it cf.}\ eqs.~(\ref{eq:flowed-R-hamiltonian}) and (\ref{eq:unflowed-R-hamiltonian}) or eqs.~(\ref{eq:flowed-NS-hamiltonian}) and (\ref{eq:unflowed-NS-hamiltonian}), the argument goes through also for the flowed sectors.

\section*{Acknowledgments}
We are grateful to Lorenz Eberhardt, Kevin Ferreira and Ben Hoare for useful conversations. This work is partially supported through a research grant of the Swiss National Science Foundation, as well as by the NCCR SwissMAP, funded by the Swiss National Science Foundation. A.S.~also acknowledges support by the ETH ``Career Seed Grant'' no.~0-20313-17. 

\appendix 

\section{Conventions}
\label{app:conventions}

For the indices we adopt the following notation. The world-sheet coordinates $\sigma$ and $\tau$ are denoted by $\alpha, \beta$; we reserve $\mu,  \nu, \rho, \dots$ for spacetime coordinate and capital letters $A, B, \dots $ for tangent space coordinates. Finally, $I, J, K \in \{ 1, 2 \}$ label rows and columns of Pauli matrices. 

Essentially following the conventions of \cite{Borsato:2015mma}, we define the three-dimensional gamma matrices for AdS$_3$ and $\rm{S}^3$ as
\begin{equation}
  \begin{aligned}
    \gamma^t &= -i\sigma_3 \ , \quad &
    \gamma^1 &= \sigma_1 \ , \quad &
    \gamma^2 &= \sigma_2 \ ,\\
    \gamma^3 &= \sigma_1 \ , \quad &
    \gamma^4 &= \sigma_2 \ , \quad &
    \gamma^\sph &= \sigma_3 \ , \\
    \gamma^5 &= \sigma_1 \ , \quad &
    \gamma^6 &= \sigma_2 \ , \quad &
    \gamma^\sphh &= \sigma_3 \ .
  \end{aligned}
\end{equation}
The resulting ten-dimensional Gamma matrices are
\begin{equation}
  \begin{aligned}
\Gamma^A & = + \sigma_1 \otimes \sigma_2 \otimes \gamma^A \otimes \mathbb{I} \otimes \mathbb{I}\ , \qquad A = t, 1 ,2\ ,  \\
\Gamma^A & = + \sigma_1 \otimes \sigma_1 \otimes \mathbb{I} \otimes \gamma^A \otimes \mathbb{I}\ , \qquad A = 3, 4, \sph\ , \\
\Gamma^A & = + \sigma_1 \otimes \sigma_3 \otimes \mathbb{I} \otimes \mathbb{I} \otimes \gamma^A\ , \qquad A = 5, 6, \sphh\ , \\ 
\Gamma^7 &= - \sigma_2 \otimes \mathbb{I} \otimes \mathbb{I} \otimes \mathbb{I} \otimes \mathbb{I}\ . 
  \end{aligned}
\end{equation}
The metric on the tangent space is 
\begin{equation}
\eta_{AB} = \rm{diag}(-1, 1, 1, 1, 1, 1, 1, 1, 1, 1)\ . 
\end{equation}
For light-cone tangent space coordinates we have 
\begin{equation}
\eta_{++} = \eta_{--} = 0\ , \qquad \eta_{+-} = 1\ , \qquad \eta^{++} = \eta^{--} = 0\ , \qquad \eta^{+-} = 1\ . 
\end{equation}

\section{Details of the Green-Schwarz action}
\label{app:details-green-schwarz-action}

The Kalb-Ramond field is given by
\begin{equation}
\begin{split}
B = & \dfrac{4 q \Rads^2}{\left(4 - X_1^2 - X_2^2 \right)^2} \left( X_1 dX_2 - X_2 dX_1 \right) \wedge dt \\
& + \dfrac{4 q R_1^2}{\left(4 + X_3^2 + X_4^2 \right)^2} \left( X_3 dX_4 - X_4 dX_3 \right) \wedge d \sph \\
& + \dfrac{4 q R_2^2}{\left(4 + X_5^2 + X_6^2 \right)^2} \left( X_5 dX_6 - X_6 dX_5 \right) \wedge d \sphh\ . 
\end{split}
\end{equation}
The fermions $\tilde{\theta}_I$ are defined as 
\begin{equation}
\begin{split}
\tilde{\theta}_1 & = \sqrt{\dfrac{1 + \tilde q}{2}} \theta_1 - \sqrt{\dfrac{1 - \tilde q}{2}} \theta_2 \ , \\
\tilde{\theta}_2 & = \sqrt{\dfrac{1 - \tilde q}{2}} \theta_1 + \sqrt{\dfrac{1 + \tilde q}{2}} \theta_2 \ ,
\end{split}
\end{equation}
where $\theta_I$ are 32-components spinors and $\tilde q = \sqrt{1 - q^2}$. Following once more the conventions of \cite{Borsato:2015mma}, the conjugation of spinors is defined as 
\begin{equation}
\bar \theta = \theta^t T\ ,  \qquad T = -i \sigma_2 \otimes \sigma_2 \otimes \sigma_2 \otimes \sigma_2 \otimes \sigma_2\ . 
\end{equation}
Moreover, 
\begin{equation}
\slashed{E}_\alpha = \partial_{\alpha} X^\mu \slashed{E}_\mu\ , \qquad \slashed{E}_\mu = \displaystyle\sum_{A = 0}^9 E_\mu^{\ A} \eta_{AB} \Gamma^B\ ,
\end{equation}
\begin{equation}
\slashed{F} = F_{ABC} \Gamma^{ABC}\ , \qquad \slashed{H}_\alpha = \partial_{\alpha}  X^\mu \slashed{H}_\mu\ , \qquad \slashed{H}_\mu = H_{\mu A B}\Gamma^{AB}\ . 
\end{equation}
Finally, the covariant derivative is defined as
\begin{equation}
D_\alpha = \partial_\alpha + \dfrac{1}{4} \partial_\alpha X^\mu \slashed{\omega}_\mu\ , \qquad \slashed{\omega}_\mu = \omega_{\mu A B} \Gamma^{AB}\ ,
\end{equation}
where $\omega$ is the spin connection. 

\section{Gauge-fixed Green-Schwarz action in plane-wave background}
\label{app:Gauge-fixed-Green-Schwarz-action-in-plane-wave-background}

For the plane-wave background defined in section \ref{sec:Penrose-limit} the vielbeins $E_\mu^{\ A}$ are
\begin{equation}
E_+^{\ +} = 1\ , \qquad E_+^{\ -} =  - \dfrac{A_{ij}x_i x_j}{2}\ , \qquad  E_-^{\ +} = 0\ , \qquad E_-^{\ -} = 1\ . 
\end{equation}
\begin{equation}
E_{i}^{\ A} = \delta_{i}^A\ , \qquad i = 1, \ldots, 8\ , \quad A = 1, \ldots, 8 \ .   
\end{equation}
They satisfy 
\begin{equation}
E_\mu^{\ A} E_\nu^{\ B} \eta_{AB} = g_{\mu \nu}\ . 
\end{equation}
The gauge fixing-choice \eqref{eq:gauge-fixing}, together with the definition of the conjugate momenta \eqref{eq:conjugate-momenta}, implies for the world-sheet metric 
\begin{equation}
\gamma^{\alpha \beta} = \begin{pmatrix} -1 & 0 \\ 0 & 1 \end{pmatrix}\ . 
\end{equation}
The fermions $\theta_I$ are defined as
\begin{equation}
    \theta_1 =
    \frac{1}{2}
    \begin{pmatrix}
      + e^{-i\pi/4} \sin\!\frac{\omega}{2} \,\theta_{14} \\ 
      + e^{-i\pi/4} \sin\!\frac{\omega}{2} \,\theta_{13} \\ 
      - e^{-i\pi/4} \cos\!\frac{\omega}{2} \,\theta_{12} \\ 
      + e^{+i\pi/4} \cos\!\frac{\omega}{2} \,\theta_{11} \\
      - e^{-i\pi/4} \cos\!\frac{\omega}{2} \,\bar{\theta}_{11} \\ 
      - e^{+i\pi/4} \cos\!\frac{\omega}{2} \,\bar{\theta}_{12} \\ 
      + e^{+i\pi/4} \sin\!\frac{\omega}{2} \,\bar{\theta}_{13} \\ 
      - e^{+i\pi/4} \sin\!\frac{\omega}{2} \,\bar{\theta}_{14}
    \end{pmatrix}
    \oplus
    \begin{pmatrix}
      + e^{-i\pi/4} \cos\!\frac{\omega}{2} \,\theta_{14} \\ 
      - e^{-i\pi/4} \cos\!\frac{\omega}{2} \,\theta_{13} \\ 
      - e^{-i\pi/4} \sin\!\frac{\omega}{2} \,\theta_{12} \\ 
      - e^{+i\pi/4} \sin\!\frac{\omega}{2} \,\theta_{11} \\
      + e^{-i\pi/4} \sin\!\frac{\omega}{2} \,\bar{\theta}_{11} \\ 
      - e^{+i\pi/4} \sin\!\frac{\omega}{2} \,\bar{\theta}_{12} \\ 
      - e^{+i\pi/4} \cos\!\frac{\omega}{2} \,\bar{\theta}_{13} \\ 
      - e^{+i\pi/4} \cos\!\frac{\omega}{2} \,\bar{\theta}_{14}
    \end{pmatrix}
    \oplus
    \begin{pmatrix}
      0 \\ 0 \\ 0 \\ 0 \\ 0 \\ 0 \\ 0 \\ 0
    \end{pmatrix}
    \oplus
    \begin{pmatrix}
      0 \\ 0 \\ 0 \\ 0 \\ 0 \\ 0 \\ 0 \\ 0
    \end{pmatrix} ,
    \label{eq:fermions1}
 \end{equation}
 \begin{equation}
    \theta_2 =
    \frac{1}{2}
    \begin{pmatrix}
      - e^{+i\pi/4} \sin\!\frac{\omega}{2} \,\theta_{24} \\ 
      - e^{+i\pi/4} \sin\!\frac{\omega}{2} \,\theta_{23} \\ 
      + e^{+i\pi/4} \cos\!\frac{\omega}{2} \,\theta_{22} \\ 
      + e^{-i\pi/4} \cos\!\frac{\omega}{2} \,\theta_{21} \\
      - e^{+i\pi/4} \cos\!\frac{\omega}{2} \,\bar{\theta}_{21} \\ 
      + e^{-i\pi/4} \cos\!\frac{\omega}{2} \,\bar{\theta}_{22} \\ 
      - e^{-i\pi/4} \sin\!\frac{\omega}{2} \,\bar{\theta}_{23} \\ 
      + e^{-i\pi/4} \sin\!\frac{\omega}{2} \,\bar{\theta}_{24}
    \end{pmatrix}
    \oplus
    \begin{pmatrix}
      - e^{+i\pi/4} \cos\!\frac{\omega}{2} \,\theta_{24} \\ 
      + e^{+i\pi/4} \cos\!\frac{\omega}{2} \,\theta_{23} \\ 
      + e^{+i\pi/4} \sin\!\frac{\omega}{2} \,\theta_{22} \\ 
      - e^{-i\pi/4} \sin\!\frac{\omega}{2} \,\theta_{21} \\
      + e^{+i\pi/4} \sin\!\frac{\omega}{2} \,\bar{\theta}_{21} \\ 
      + e^{-i\pi/4} \sin\!\frac{\omega}{2} \,\bar{\theta}_{22} \\ 
      + e^{-i\pi/4} \cos\!\frac{\omega}{2} \,\bar{\theta}_{23} \\ 
      + e^{-i\pi/4} \cos\!\frac{\omega}{2} \,\bar{\theta}_{24}
    \end{pmatrix}
    \oplus
    \begin{pmatrix}
      0 \\ 0 \\ 0 \\ 0 \\ 0 \\ 0 \\ 0 \\ 0
    \end{pmatrix}
    \oplus
    \begin{pmatrix}
      0 \\ 0 \\ 0 \\ 0 \\ 0 \\ 0 \\ 0 \\ 0
    \end{pmatrix} ,
  \label{eq:fermions2}
\end{equation}
where $\bar \theta_{Ii}$ is the complex conjugate of $\theta_{Ii}$ for $i = 1, \ldots, 4$. One can easily verify that eqs.\ \eqref{eq:fermions1} and \eqref{eq:fermions2} are in agreement with the choice of kappa symmetry fixing \eqref{eq:gauge-fixing}. 

\subsection{Bosonic Hamiltonian}

Exploiting \eqref{eq:H=-p_+} the explicit form of the light-cone bosonic Hamiltonian can be found by solving the quadratic equation $C_2 = 0$, where $C_2$ has been defined in \eqref{eq:virasoro-constraints}. In the chosen gauge-fixing \eqref{eq:gauge-fixing}, for the plane-wave background we find
\begin{equation}
\mathcal{H}_b = \frac{1}{2} p_i p_i + \frac{1}{2} \pri x_i \pri x_i + \frac{1}{2} A_{ij} x_i x_j + B_{+i}\pri x_i \ . 
\end{equation}
Since the only non-vanishing components of the Kalb-Ramond field are
\begin{equation}
\begin{aligned}
B_{+1} & = q x_2\ , & B_{+2} & = -q x_1\ , \\
B_{+3} & = q  x_4\, \cos \varphi\,  \cos \omega\ , & B_{+4} & = - q x_3\, \cos \varphi \, \cos \omega\ , \\
B_{+5} & = q x_6\, \sin \varphi \, \sin \omega \ , & B_{+6} & = - q x_5 \, \sin \varphi \, \sin \omega \ , 
\end{aligned}
\end{equation}
we recover \eqref{eq:bosonic-hamiltonian}. The complex coordinates $\textrm{x}_i,  \bar{\textrm{x}}_i$ with $i = 1, \dots, 4$ have been (implicitly) introduced in subsection \ref{subsec:mode-expansion-and-normal-ordering} and can be expanded in modes according to
\begin{align}
\textrm{x}_i = & \dfrac{1}{\sqrt{r}} \displaystyle\sum_{n \in \mathbb{Z}} \left( \dfrac{a_n^{i, + \dagger}}{\sqrt{\omega_n^{i,+}}} e^{i \omega_n^+ \tau - i \frac{2 \pi n}{r} \sigma} + \dfrac{a_n^{i,-}}{\sqrt{\omega_n^-}} e^{-i \omega_n^- \tau +i \frac{2 \pi n}{r} \sigma} \right),  \\
p_\textrm{x}^i = & \tfrac{1}{2} \dot{\bar{\textrm{x}}}^i =  \dfrac{i}{2\sqrt{r}} \displaystyle\sum_{n \in \mathbb{Z}} \left( \sqrt{\omega_n^-} a_n^{i,- \dagger} e^{i \omega_n^- \tau -i \frac{2 \pi n}{r} \sigma} - \sqrt{\omega_n^+} a_n^{i,+} e^{-i \omega_n^+ \tau +i \frac{2 \pi n}{r} \sigma} \right), 
\end{align}
where we have used the short-hand notation $\omega_n^\pm = \omega_n^\pm(\mu_i^b)$. The fields $\textrm{x}_i$ obey periodic boundary conditions 
\begin{equation}
\textrm{x}_i(\tau, \sigma) = \textrm{x}_i(\tau, \sigma + r)
\end{equation}
and canonical commutation relations
\begin{equation}
[\textrm{x}_i (\tau, \sigma), p_\textrm{x}^j(\tau, \sigma')] = i \delta_i^j \delta(\sigma - \sigma')\,,
\end{equation}
that are in turn equivalent to \eqref{eq:bosonic-commutation-relations}. 

\subsection{Fermionic Hamiltonian}

The expressions in \eqref{eq:GS-L-kin} and \eqref{eq:GS-L-WZ} simplify drastically  for the choice of gauge and kappa symmetry made in \eqref{eq:gauge-fixing}. In fact, we find 
\begin{equation}
\slashed{E}_\tau = \Gamma^- + \left( \dot x^- - \tfrac{1}{2} A_{ij}x_i x_j \right) \Gamma^+ + \dot x_i \Gamma^i,  
\end{equation}
\begin{equation}
\slashed{E}_\sigma = \pri x^- \Gamma^+ + \pri x_i \Gamma^i, 
\end{equation}
\begin{equation}
\slashed{\omega}_\mu \propto \Gamma^+. 
\end{equation}
Exploiting \eqref{eq:gamma+-identities} we obtain
\begin{equation}
\begin{split}
\mathcal{L}_{kin} & = i \bar{ \tilde \theta}_I \Gamma^\perp \Gamma^-(\tilde q \sigma_3^{IJ} + q \sigma_1^{IJ})\tilde \theta_J + i \bar{ \tilde \theta}_I \Gamma^- \partial_\tau \tilde \theta_I \\
& = \frac{i}{2} \bar \theta_I \Gamma^\perp \Gamma^- \sigma_3^{IJ} \theta_J + i \bar \theta_I \Gamma^- \dot \theta_I\ ,  
\end{split}
\end{equation}
where 
\begin{equation}
\Gamma^\perp = \Gamma^{12} + (\cos \varphi \cos \omega)\,  \Gamma^{34} + (\sin \varphi \sin \omega)\,\Gamma^{56} \ ,
\end{equation}
and 
\begin{equation}
\mathcal{L}_{WZ} = i \bar{\tilde \theta}_I \sigma_1^{IJ} \Gamma^- \partial_\sigma \tilde \theta_J = i q \bar \theta_I \sigma_3^{IJ} \Gamma^- \pri \theta_J + i \tilde q \bar \theta_I \sigma_1^{IJ} \Gamma^- \pri \theta_J\ .
\end{equation}
According to the definitions \eqref{eq:gamma+-}, \eqref{eq:fermions1} and \eqref{eq:fermions2} we find that $\mathcal{L}_f = \mathcal{L}_{kin} + \mathcal{L}_{WZ} $ reduces to 
\begin{equation}
\mathcal{L}_f =  \sum_{i=1}^4 [i \bar{\theta}_{1i} ( \dot{\theta}_{1i} - i \tilde q \pri{\theta}_{2i} + q \pri{\theta}_{1i} )
+ i \bar{\theta}_{2i} ( \dot{\theta}_{2i} + i \tilde q \pri{\theta}_{1i} - q \pri{\theta}_{2i} ) ] -\sum_{i=1}^4 \mu_i^f \left( \bar{\theta}_{1i} \theta_{1i}  - \bar{\theta}_{2i} \theta_{2i} \right)\ ,   
\end{equation}
where $\mu_1^f, \mu_2^f, \mu_3^f, \mu_4^f$ have been defined in \eqref{eq:fermion-masses-GS}. The fermionic fields $\theta_{1i}, \theta_{2i}$ can be expanded in modes following, \textit{e.g.}~\cite{Lloyd:2014bsa}
\begin{equation}
\begin{gathered}
\theta_{1i} = \dfrac{e^{-i \tfrac{\pi}{4}}}{\sqrt{r}} \displaystyle\sum_{n \in \mathbb{Z}} \left( \dfrac{g_n^-}{\sqrt{\omega_n^-}} b_n^{i, - \dagger} e^{i \omega_n^- \tau - i \frac{2 \pi n}{r}  \sigma} - \dfrac{f_n^+}{\sqrt{\omega_n^+}} b_n^{i,+} e^{-i \omega_n^+ \tau + i \frac{2 \pi n}{r}  \sigma}  \right)\ ,  \\
\theta_{2i} = \dfrac{e^{i \tfrac{\pi}{4}}}{\sqrt{r}} \displaystyle\sum_{n \in \mathbb{Z}} \left( \dfrac{g_n^+}{\sqrt{\omega_n^+}} b_n^{i,+} e^{-i \omega_n^+ \tau + i \frac{2 \pi n}{r}  \sigma} - \dfrac{f_n^-}{\sqrt{\omega_n^-}} b_n^{i,- \dagger} e^{i \omega_n^- \tau - i \frac{2 \pi n}{r}  \sigma}  \right)\ , 
\end{gathered}
\end{equation}
where we have used the short-hand notation $\omega_n^\pm = \omega_n^\pm(\mu_i^f)$ and similarly for $f_n^\pm$ and $g_n^\pm$ defined as
\begin{equation}
f_n^\pm = \sqrt{\dfrac{\mu \pm q \frac{2 \pi n}{r}  + \omega_n^\pm}{2}}\ , \qquad g_n^\pm = - \dfrac{\pi \tilde q n}{r  f_n^\pm}\ . 
\label{eq:f_n-g_n-definition}
\end{equation}
They satisfy 
\begin{equation}
f_{-n}^\pm = f_n^\mp\ , \qquad g_{-n}^\pm = -g_n^\mp\ , \qquad (f^\pm_n)^2 + (g^\pm_n)^2 = \omega_n^\pm\ . 
\end{equation}
Note that for pure NS-NS flux \eqref{eq:f_n-g_n-definition} reduces to
\begin{equation}
\begin{aligned}
& f_n^+ = \sqrt{\left|\frac{2 \pi n}{r}  + \mu \right|} \Theta\left(\frac{2 \pi n}{r}  + \mu\right)\ ,  &  & f_n^- = \sqrt{\left|\frac{2 \pi n}{r}  - \mu\right|} \Theta\left(\frac{2 \pi n}{r}  - \mu\right)\ ,   \\
& g_n^+ = \sqrt{\left|\frac{2 \pi n}{r}  + \mu\right|} \Theta\left(-\frac{2 \pi n}{r}  - \mu \right)\ ,  &  & g_n^- = \sqrt{\left|\frac{2 \pi n}{r}  - \mu\right|} \Theta\left(-\frac{2 \pi n}{r}  + \mu\right)\ ,   
\end{aligned}
\end{equation}
where $\Theta(x)$ is the Heaviside step function. Finally, the fermionic fields obey the anticommutation relations
\begin{equation}
\begin{array}{l}
\{ \theta_{1i}(\tau, \sigma), \bar \theta_{1j}(\tau, \sigma') \} = \delta_{ij} \delta(\sigma - \sigma')\ , \\
\{ \theta_{2i}(\tau, \sigma), \bar \theta_{2j}(\tau, \sigma') \} = \delta_{ij} \delta(\sigma - \sigma')\ ,
\end{array}
\end{equation}
that are equivalent to \eqref{eq:fermionic-commutation-relations}. 

\section{Superconformal affine algebras and spectral flow automorphism}
\label{app:superconformal-affine-algebras-and-spectral-flow-automorphism}

The $\mathcal{N} = 1$ affine algebra $\mathfrak{sl}(2)^{(1)}_k$ is characterised by the commutation relations
\begin{equation}
\begin{aligned}
& [\mathcal{K}^+_m, \mathcal{K}^-_n] = - 2 \mathcal{K}_{m+n}^3 + k m \delta_{m,-n} &  & [\mathcal{K}^3_m, \mathcal{K}^\pm_n] = \pm \mathcal{K}^\pm_{m+n} & & [\mathcal{K}^3_m, \mathcal{K}^3_n] = -\frac{k}{2} m \delta_{m, -n}  \\
& [\mathcal{K}_m^\pm, \psi_r^3] = \mp \psi_{r+m}^\pm & & [\mathcal{K}_m^3, \psi_r^\pm] = \pm \psi_{r+m}^\pm & & [\mathcal{K}_m^\pm, \psi_r^\mp] = \mp 2 \psi^3_{m+r} \\
& \{ \psi^+_r, \psi^-_s \} = k \delta_{r, -s} & & \{ \psi^3_r, \psi^3_s \} = -\frac{k}{2} \delta_{r, -s}\ . 
\end{aligned}
\label{eq:sl(2)-kac-moody-algebra}
\end{equation}
Similarly, for the $\mathfrak{su}(2)_{k'}$ superconformal affine algebra we have 
\begin{equation}
\begin{aligned}
& [\mathcal{J}^+_m, \mathcal{J}^-_n] = 2 \mathcal{J}_{m+n}^3 + k' m \delta_{m,-n} &  & [\mathcal{J}^3_m, \mathcal{J}^\pm_n] = \pm \mathcal{J}^\pm_{m+n} & & [\mathcal{J}^3_m, \mathcal{J}^3_n] = \frac{k'}{2} m \delta_{m, -n}  \\
& [\mathcal{J}_m^\pm, \chi_r^3] = \mp \chi_{r+m}^\pm & & [\mathcal{J}_m^3, \chi_r^\pm] = \pm \chi_{r+m}^\pm & & [\mathcal{J}_m^\pm, \chi_r^\mp] = \pm 2 \chi^3_{m+r} \\
& \{ \chi^+_r, \chi^-_s \} = k' \delta_{r, -s} & & \{ \chi^3_r, \chi^3_s \} = \frac{k'}{2} \delta_{r, -s}\ . 
\end{aligned}\label{eq:su(2)-kac-moody-algebra}
\end{equation}
In each case one can define decoupled bosonic currents commuting with the fermions. These decoupled currents then have levels $k$ and $k'$ that  are shifted by $+2$ and $-2$, respectively, see \cite{Ferreira:2017pgt} for a detailed description of this construction. The ${\cal N}=1$ supersymmetric affine algebra is in fact ${\cal N}=1$ superconformal since the generators of the ${\cal N}=1$ superconformal algebra can be constructed out of the currents and fermions. (For the stress-energy tensor this is just the Sugawara construction applied to the decoupled bosonic currents together with the usual free field formula for the fermions, see {\it e.g.}\ \cite{Ferreira:2017pgt} for a detailed exposition.)

For any $w,w'\in\mathbb{Z}$, the spectral flow automorphism of $\mathfrak{sl}(2)^{(1)}_k$ and $\mathfrak{su}(2)_{k'}$  is defined by replacing the original unhatted generators by the hatted expressions defined by
\begin{equation}
\begin{aligned}
& \hat{\mathcal{K}}_n^\pm = \mathcal{K}^\pm_{n \mp w}\ , & \qquad  & \hat{\mathcal{J}}_n^\pm = \mathcal{J}^\pm_{n \pm w'}\ , \\
& \hat{\mathcal{K}}_n^3 = \mathcal{K}^3_n + \frac{k}{2}w \, \delta_{n,0}\ , & \qquad & \hat{\mathcal{J}}_n^3 = \mathcal{J}^3_n + \frac{k'}{2}w' \, \delta_{n,0}\ ,  \\
& \hat{L}_n^{\mathfrak{sl}} = L_n^{\mathfrak{sl}} - w \, \mathcal{K}_n^3 - \dfrac{k}{4}w^2 \delta_{n,0}\ , & \qquad & \hat{L}_n^{\mathfrak{su}} = L_n^{\mathfrak{su}} + w' \, \mathcal{J}_n^3 + \dfrac{k'}{4}{w'}^2 \delta_{n,0}\ , \\
& \hat \psi_r^3 = \psi_r^3\ , & & \hat \chi_r^3 = \chi_r^3\ , \\
& \hat \psi_r^\pm = \psi^\pm_{r \mp w}\ ,   & & \hat \chi_r^\pm = \chi^\pm_{r \pm w'}\ .  
\end{aligned}
\label{eq:spectrally-flowed-generators}
\end{equation}
It is straightforward to verify that this map defines indeed an automorphism of the two superconformal affine algebras, {\it i.e.}\ that the hatted generators satisfy the same commutation and anti-commutation relations as the original generators.

\bibliographystyle{JHEP}
\bibliography{refs}

\makeatletter \@ifundefined{Sphere}{\newcommand{\Sphere}{\text{S}}}{}
  \@ifundefined{AdS}{\newcommand{\AdS}{\text{AdS}}}{}
  \@ifundefined{CFT}{\newcommand{\CFT}{\text{CFT}}}{}
  \@ifundefined{CP}{\newcommand{\CP}{\text{CP}}}{}
  \@ifundefined{Torus}{\newcommand{\Torus}{\text{T}}}{}
  \@ifundefined{superN}{\newcommand{\superN}{\mathcal{N}}}{}
  \@ifundefined{grpOSp}{\newcommand{\grpOSp}{\mathrm{OSp}}}{}
  \@ifundefined{grpPSU}{\newcommand{\grpPSU}{\mathrm{PSU}}}{}
  \@ifundefined{grpSU}{\newcommand{\grpSU}{\mathrm{SU}}}{}
  \@ifundefined{grpU}{\newcommand{\grpU}{\mathrm{U}}}{}
  \@ifundefined{grpD}{\newcommand{\grpD}{\mathrm{D}}}{}
  \@ifundefined{grpSL}{\newcommand{\grpSL}{\mathrm{SL}}}{}
  \@ifundefined{grpSp}{\newcommand{\grpSp}{\mathrm{Sp}}}{}
  \@ifundefined{grpUSp}{\newcommand{\grpUSp}{\mathrm{USp}}}{}
  \@ifundefined{grpSO}{\newcommand{\grpSO}{\mathrm{SO}}}{}
  \@ifundefined{grpO}{\newcommand{\grpO}{\mathrm{O}}}{}
  \@ifundefined{algOSp}{\newcommand{\algOSp}{\mathfrak{osp}}}{}
  \@ifundefined{algPSU}{\newcommand{\algPSU}{\mathfrak{psu}}}{}
  \@ifundefined{algSU}{\newcommand{\algSU}{\mathfrak{su}}}{}
  \@ifundefined{algSp}{\newcommand{\algSp}{\mathfrak{sp}}}{}
  \@ifundefined{algSL}{\newcommand{\algSL}{\mathfrak{sl}}}{}
  \@ifundefined{algGL}{\newcommand{\algGL}{\mathfrak{gl}}}{}
  \@ifundefined{algU}{\newcommand{\algU}{\mathfrak{u}}}{}
  \@ifundefined{algSO}{\newcommand{\algSO}{\mathfrak{so}}}{}
  \@ifundefined{algO}{\newcommand{\algO}{\mathfrak{o}}}{}
  \@ifundefined{Integers}{\newcommand{\Integers}{\mathbb{Z}}}{}
  \@ifundefined{Reals}{\newcommand{\Reals}{\mathbb{R}}}{} \makeatother

\providecommand{\href}[2]{#2}\begingroup\raggedright\endgroup

\end{document}